\documentclass[10pt,journal,compsoc]{IEEEtran}
\ifCLASSOPTIONcompsoc
  \usepackage[nocompress]{cite}
\else
  \usepackage{cite}
\fi
\usepackage{graphicx}  
\usepackage{array} 
\usepackage{epstopdf}
\usepackage{siunitx}
\usepackage{makecell, booktabs, caption}
\usepackage{amsmath}
\usepackage{amssymb}
\ifCLASSINFOpdf
\else
\fi
\begin{document}

%
\title{PSAA: Provable Secure and Anti-Quantum Authentication Based on Randomized RLWE for Space Information Network}
%
%


\author{Junyan~Guo,
       Ye~Du,
       Xuesong~Wu, 
        Meihong~Li,
        Runfang~Wu,
        Zhichao~Sun
\IEEEcompsocitemizethanks{\IEEEcompsocthanksitem J. Guo, Y. Du, X. Wu, M. Li, R. Wu and Z. Sun are with the Beijing Key Laboratory of Security and Privacy in Intelligent Transportation, Beijing Jiaotong University, Beijing 100044, China (Corresponding author: Y. Du, e-mail: ydu@bjtu.edu.cn).\protect\\
}

}

%
%

\markboth{Journal of \LaTeX\ Class Files,}%
{Shell \MakeLowercase{\textit{et al.}}: Bare Demo of IEEEtran.cls for Computer Society Journals}
%



\IEEEtitleabstractindextext{%
\begin{abstract}
	Currently, due to the high scalability and global coverage of space information network (SIN), more service providers and users are willing to provide or subscribe to personal services through the satellite network. However, the messages are transmitted in public satellite-ground links, which makes access users vulnerable to various forms of attacks. Existing authentication protocols do not meet the expected security and short delay requirements to ensure the security of real-time user access and the confidentiality of communication content. 
	Moreover, with the development of quantum computers, the difficult problems such as ECDLP and DLP have also been proven to be solvable in polynomial time, leading to new threats. Therefore, in this paper, we propose a provably secure and anti-quantum authentication protocol based on randomized RLWE. The protocol not only meets the pre-defined security requirements, but also reduces the total delay of the authentication phase based on the pre-negotiation and fewer authentication transmission. In addition, a concise handover scheme is designed for signal handover scenarios caused by satellite dynamic topology. Further rigorous formal and informal security proofs and performance analysis show that our proposed protocol is more applicable to SIN, while ensuring higher security and resisting various attacks with lower authentication delay.

\end{abstract}

}

\maketitle

\IEEEdisplaynontitleabstractindextext

%
\IEEEpeerreviewmaketitle

\IEEEraisesectionheading{\section{Introduction}\label{sec:introduction}}

%
%
%
%

\IEEEPARstart{W}{ith} the continuous development of satellite communication technology, global services free from time and geography constraints will gradually become possible \cite{SpaceTerrestrialNetwork}. 
Space information network (SIN) as a large integrated heterogeneous network, which includes satellite backbone networks, inter-satellite links and satellite-ground links in a broad sense, is proposed in this context to meet the needs of land-based, sea-based and space-based users for real-time communication, navigation, disaster warning and other services. Compared with traditional terrestrial networks, SIN has unique advantages. First, as an extension of the terrestrial network, SIN provides reliable communication and broadband services through satellite networks to sparsely populated and remote areas such as the oceans, deserts, or mountainous regions at a lower cost \cite{GuoFog}. Secondly, as long as it has the ability to access and communicate with satellites, regardless of geographic location, service providers anywhere can provide subscription services to users around the world. In addition, through the unified acquisition, coordination and integration of satellite resources and ground resources, SIN can more easily meet the needs of users for personalized and diverse \cite{ChaoEmerging}.
 
At present, the Internet of Things (IoT) technology is booming, and the IoT devices have been applied in various areas of work and life, such as smart home, intelligent transportation and industrial control systems \cite{Xuehandover}. In some scenarios, IoT devices are distributed in areas that cannot be covered by traditional terrestrial networks and can only access SIN. 
However, due to the openness of the satellite-to-ground link, the devices are vulnerable to various attacks initiated by the adversary, such as eavesdropping attacks, replay attacks, impersonation attacks, man-in-the-middle attacks, device loss attacks and insider attacks, when they access to the network and then transmit sensitive data \cite{JiangSecurity}. 
The authentication protocol is an effective measure to verify the legitimacy of access users and prevent malicious nodes from accessing and performing destructive behaviors.
If no effective access authentication protocol is adopted, not only device privacy and confidential data will be leaked, but also the computing and communication resources of SIN will be greatly consumed, which will have a huge negative impact on the stability of the whole society. 
Although many authentication protocols have been proposed by scholars, they are not suitable for providing secure and reliable communication services to time-sensitive users because of security vulnerabilities or long authentication delay. 
Not only the above problems, most existing protocols are proposed based on public key cryptography mechanisms such as the discrete logarithm problem (DLP) and the elliptic curve discrete logarithm problem (ECDLP), and these difficult assumptions have been proved to be solved by Shor algorithm \cite{ShorAlgorithms} in polynomial time, which ultimately makes the protocols vulnerable. 
In addition, although Ma et al. \cite{ma2019laa} proposed the first lattice-based anti-quantum three-party authentication protocol, the security drawbacks and extremely large communication overhead make this protocol not applicable to SIN.
Therefore, in this paper, we propose a sufficiently secure and low delay authentication protocol for SIN that supports mutual authentication and negotiation the session key for SIN. 
In particular, our main contributions in this paper are as follows: 

\begin{itemize}
	\item [(1)]
	We propose a new provable secure and anti-quantum authentication protocol for SIN, named PSAA protocol. 
	In the authentication phase, the satellite replaces the terrestrial control station to verify the legitimate identity of the access user, and as a relay node enable the terrestrial control station and the user to negotiate a secure session key, which reduces the traditional at least 4 satellite-to-ground transmissions to only 2 times. 
	Moreover, for the signal handover scenario caused by the satellite mobile topology, we also designed a low computing and storage load handover verification scheme to enable the next satellite and the user to mutually verify the legitimacy of each other.
	\item [(2)]
	In the PSAA protocol, we add unpredictable random sample values based on randomized ring learning with errors (RLWE) to avoid key reuse attacks without affecting mutual authentication, key negotiation and identity privacy.  
	
	\item [(3)]
	The security of PSAA protocol is proved by two formal and the informal analysis methods, which proves not only the semantic security of the negotiated session key, but also meets the security requirements and resistance to various forms of attacks.
	

	\item [(4)]
	The performance analysis and comparison of security attributes, authentication delay and communication overhead with related protocols show that the PSAA protocol has less authentication delay and communication overhead than another anti-quantum protocol and is more suitable for the user to access SIN under the premise of ensuring higher security and quantum resistance.
\end{itemize}

The rest of the paper is organized as follows. Section \ref{RelatedWork} involves the related work. Section \ref{MathematicalPreliminaries} introduces the mathematical preliminaries. Section \ref{Model} details the system model, threat model and security requirements. Section \ref{ProposedProtocol}  elaborates our proposed PSAA protocol. Sections \ref{SecurityAnalysis} and \ref{PerformanceComparison} analyze the security and performance, respectively. Finally, Section \ref{Conclusion} draws the conclusion of this paper.


\section{Related Work} \label{RelatedWork}

In recent years, researchers have successively proposed many authentication protocols for SIN. Cruickshank et al. \cite{Cruickshank1996} first proposed the prototype of the authentication protocol. However, in the protocol \cite{Cruickshank1996}, the authentication messages between the user and the terrestrial control station are all encrypted and decrypted by the public key cryptographic mechanism, which makes the system to bear the complex computation overhead. In addition, user privacy is also not properly protected. 
To overcome the shortcomings in \cite{Cruickshank1996}, Hwang et al. \cite{Hwang2003} proposed using symmetric encryption mechanism to reduce the computational overhead during authentication process and to protect the true identity of user with a temporary identity, which has also been shown to be resistant to replay attacks. 
However, Chang et al. \cite{Chang2005} showed that \cite{Hwang2003} does not satisfy forward security, and if the session key is leaked, the key security of other sessions will be affected.  
Since then, Chen et al. \cite{Chen2009} proposed an authentication protocol that does not need to store the certificate and can be self-verified based on the message authentication code (MAC). 
In 2012, Zheng et al. \cite{Zheng2012} proposed a lightweight authentication protocol based only on hash functions and XOR operations. Besides, \cite{Zheng2012} was proved to be mutually authenticated by SVO logic. However, although \cite{Zheng2012} greatly reduces the computational overhead, the high authentication delay caused by 9 message transmissions is not suitable for real-time communication. In addition, Zhao et al. \cite{Zhao2016} found \cite{Zheng2012} vulnerable to identity spoofing attacks. 

In 2019, Ostad-Sharif et al. \cite{Ostad-Sharif} and Qi et al. \cite{QiECCsatellite} both proposed authentication and session key negotiation protocols based on elliptic curve cryptography.
In these two protocols, the satellite acts as a relay node between the user and the terrestrial control station during the authentication phase, only forwarding the authentication message without verifying the validity of the message and the sender. As a result, the adversary can send a large number of useless messages to the satellite to consume the precious satellite resources. 
In addition, \cite{Ostad-Sharif} implements three-factor verification including smart device, password, and biometric when the user logs in to avoid device loss attacks,
while \cite{QiECCsatellite} only verifies the two-factor, making it more vulnerable to offline password attacks and smart device loss attacks. For \cite{Ostad-Sharif} and \cite{QiECCsatellite}, the authentication is completed at least 4 satellite-to-ground transmissions, which also causes the authentication delay to be too long to be suitable for time-sensitive users. Recently, Yang et al. \cite{AnFRA} and Xue et al. \cite{Xuehandover} proposed that satellites verify the legitimacy of access users without the need for TCS to be responsible for access control, which reduces the authentication delay. However, neither of these two protocols can resist device loss attacks, that is, the adversary can access SIN and obtain network services after obtaining the device lost by the legitimate user. Besides, with the development of quantum computers, the difficult problems based on elliptic curve cryptography have been proved to be solved in polynomial time \cite{ShorAlgorithms}, which makes the protocols \cite{Ostad-Sharif,QiECCsatellite,AnFRA,Xuehandover} no longer secure in the post-quantum era.

Recently, Ma et al. \cite{ma2019laa} proposed anti-quantum access authentication scheme based on lattice and Gentry’s encryption mechanism. 
However, this scheme does not design a specific mechanism to resist replay attacks and device loss attacks. Moreover, the longer authentication delay and extremely large communication overhead make \cite{ma2019laa} unsuitable for resource-constrained users and SIN.

\section{Mathematical Preliminaries} \label{MathematicalPreliminaries}
In this section, we briefly review and introduce the mathematical preliminaries of ring learning with errors (RLWE) and biometric fuzzy extractor involved in our proposed protocol.
\subsection{Ring Learning With Errors}
Define the quotient ring of a polynomial $R_q=Z_{q}[x]/(x^n+1)$, where $n$ is a power of 2 and $q$ is a prime number with modulo $n$ equal to 1. The coefficient vector of any polynomial element $a$ in $R_q$ can be expressed as $a=(a_0,a_1,\dots,a_{n-1})$. The Gaussian distribution on $R_q$ is defined as $\chi_{\beta}$, where the positive real number $\beta$ is the standard deviation of the distribution.

Let $Z_q=\{-\frac{q-1}{2},\dots,\frac{q-1}{2}\}$ and the middle set $E=\{-\lfloor \frac{q}{4}\rfloor,\dots,\lfloor\frac{q}{4}\rfloor\}$, then the error reconciliation of RLWE has the following two core functions, namely $Cha$ and $Mod_2$. $Cha$ is considered a characteristic function, that is, if $x \in E$ then $Cha(x)=1$, otherwise $Cha(x)=0$. The other is the auxiliary modular function, $Mod_2(v,w)=(v+w\cdot\frac{q-1}{2})\bmod q \bmod 2$, where $v\in Z_q$, $w=Cha(v)$ \cite{QiFeng,DingAsimple}. These two functions have the following lemma.

\textbf{Lemma 1}: Given an odd prime number $q$ and two elements $v, w \in Z_q$, then $Mod_2(v,Cha(v))=Mod_2(w,Cha(v))$, where $\left| v-w\right|\leq \frac{q}{4}-2$. Similarly, this lemma also holds for the polynomial ring elements in $R_q$ \cite{DingKeyexchange}. 

Since the key reuse attack has been proved to have a security threat to the original reconciliation-based RLWE scheme, that is, using the same public key more than the usage threshold will cause the adversary to recover the corresponding private key through $Cha$ function values \cite{Dingattacforreuse}. In order to fix this vulnerability, \cite{Gaorandomized} proposed a randomized RLWE scheme, which improves randomization by adding additional unpredictable error values to prevent the adversary from obtaining the private key. Our proposed PSAA protocol is based on the randomized RLWE of \cite{Gaorandomized}, which avoids the threat of private key leakage caused by using the same public key.

\textbf{Definition 1}: RLWE assumption: Let $R_q$ and $\chi_{\beta}$ be as defined above and a fix polynomial ring elements $s\in R_q$, $A_{s,\chi_\beta}$ is the distribution over $(a, a\cdot s+2\cdot e)$, where $a$, $e$ are randomly sampled from $R_q$. RLWE assumption implies that it is difficult to distinguish $A_{s,\chi_\beta}$ from uniform distribution on $R^2_{q}$ in polynomial time, and it can be reduced to Shortest Vector Problem (SVP) in the lattice \cite{DingAsimple,DingKeyexchange,Gaorandomized}.

\subsection{Biometric Fuzzy Extractor}
The biometric key is more secure than the password in that it has higher entropy and is not only able to identify users with higher accuracy but also can't be forgotten \cite{Guopeer}. The biometric fuzzy extractor can generate a biometric key based on biometric features such as iris, fingerprint, etc., and re-extract the same biometric key from the biometric features within the error threshold during re-identification \cite{DodisFuzzy}. The fuzzy extractor contains two functions, $Gen$ and $Rep$, which are described in detail as follows. 

\begin{itemize}
	\item
	$Gen(BIO)$ is the probabilistic generation function that inputs the biometric $BIO$ and outputs the biometric key $\sigma$ and the public auxiliary data $v$.
	\item
	$Rep(BIO^*,v)$ is the reproduction function that retrieves the biometric key $\sigma$ from the auxiliary data $v$ and the re-entered biometric $BIO^*$ if and only if the error between $BIO$ and $BIO^*$ is less than the threshold $\delta$.
\end{itemize}

	

\section{System Model, Threat Model and Security Requirements} \label{Model}
This section describes in detail the system model, threat model and security requirements in SIN.

\subsection{System Model}

\begin{figure}[ht] 
	\centering  
	\includegraphics[width=0.5\textwidth]{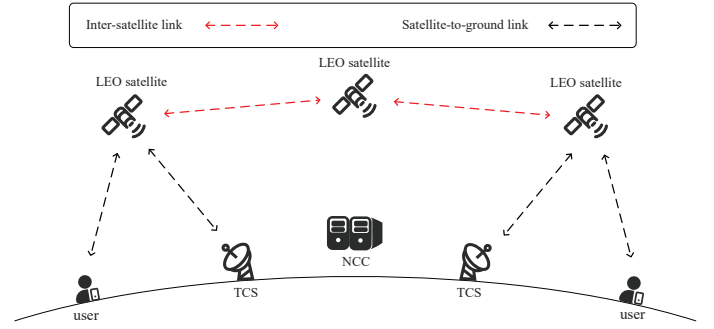}  
	\caption{System model}  
	\label{systemmodel}  
\end{figure}

As shown in Figure \ref{systemmodel}, the SIN traditionally contains four types of entities: network control center (NCC), terrestrial control station (TCS), satellite node and user. The following details
describe the functions of each entity. 

\begin{itemize}
	\item
	As mentioned in \cite{AnFRA,Xuehandover}, NCC is a fully trusted control and monitoring center for the entire system, providing registration services for satellite nodes and terrestrial control stations, and has the highest level of security protection so that no other adversary can break its security mechanism and obtain confidential information. 
	\item
	TCS provides registration services for users and subscription services for registered users through relay satellites. 
	\item
	Satellite node serves as a relay node for communication between the user and the TCS, and the subscription service provided by the TCS can be forwarded to the user through the satellite node. To ensure the signal strength and the quality of service for time-sensitive users, low earth orbit (LEO) satellites that is closer to the user are usually used as relay nodes \cite{Ostad-Sharif}. In addition, the satellite already has the ability to perform complex calculations \cite{AnFRA,Xuehandover}.
	\item
	User registered at TCS during the registration phase can use the smart device to calculate, store and communicate with satellites to receive subscription services provided by TCS. 
\end{itemize}

\subsection{Threat Model}
The threat model of the proposed protocol defines the adversary’s capabilities based on the widely used Dolev-Yao model \cite{Dolevyaomodel} and Canetti-Krawczyk model \cite{CKmodel}, that is, adversaries can not only control insecure or public channels in SIN, but also obtain the confidential information in the user's smart device. The adversary assumes the following specific capabilities. 

\begin{itemize}
	\item
	Adversary can eavesdrop, store and replay all the authentication interaction information in the public channel.
	\item
	Adversary can forge or modify the eavesdropped authentication information and resend it to the legitimate node.
	\item
	Adversary can obtain the information stored in the user device through power analysis  \cite{Poweranalysis}.
	\item
	Adversary knows the details of the authentication protocol and can obtain a clear and correct response after sending an authentication message as a participant of the protocol.

\end{itemize}

\subsection{Security Requirements}
Based on existing authentication protocols, we conclude that a well-designed and secure protocol in SIN should meet the following security requirements.

\begin{itemize}
	\item
	Mutual authentication. When the user accesses SIN, the user and the satellite should be able to mutually verify the legitimacy of each other.
	\item
	Key negotiation. During the authentication phase, the user and TCS should be able to negotiate a secure session key to protect the next communication, and the session key should be determined by both parties, not only by one party.
	\item
	Identity anonymous. The user's identity is sensitive and private information that should be anonymized to prevent the adversary from obtaining its true identity. 
	\item
	Perfect forward/backward secrecy. The security of previous and future session keys is not compromised when the current session key is obtained by the adversary.
	\item
	Attacks resistance. To ensure the reliability and security of the authentication process, the protocol proposed should be able to withstand various forms of attacks such as eavesdropping attacks, replay attacks, impersonation attacks, man-in-the-middle attacks, device loss attacks, and insider attacks. In addition, because the rapid development of quantum computing poses a great threat to authentication protocols based on elliptic curve and discrete logarithmic cryptography, our proposed protocol should also be resistant to quantum attacks to avoid leakage of privacy and confidential information.
	
\end{itemize}


\section{Proposed Protocol: PSAA} \label{ProposedProtocol}
In this section, we describe in detail the proposed PSAA protocol in the order of the system initialization phase, registration phase, pre-negotiation phase, login and authentication phase, handover phase, password and biometric update phase. In addition, to introduce the PSAA protocol more conveniently and unambiguously, we simplify the system to include only one satellite $L_j$, one TCS and one user $u_i$. The symbols involved in the protocol are shown in Table \ref{Symbol list}.

\begin{table}[htp]
	\caption{Symbol list}
	\centering
	\begin{tabular}{p{2cm} <{}  p{5.5cm} <{} }
		\hline
		\specialrule{0em}{1pt}{1pt}
		Notation      	& 	Description   \\
		\hline
		\specialrule{0em}{1pt}{1pt}
		$ID_j, ID_{tcs}$	&	Identity of $L_j$ and TCS		\\ \specialrule{0em}{1pt}{1pt}
		$ID_i$			&	True identity of $u_i$			\\ \specialrule{0em}{1pt}{1pt}
		$TID_i$			&	Temporary Identity of $u_i$		\\ \specialrule{0em}{1pt}{1pt}
		$PW_i, BIO_i$	&	Password and biometric of $u_i$\\ \specialrule{0em}{1pt}{1pt}
		$q$				&	Odd prime						\\ \specialrule{0em}{1pt}{1pt}
		$n$				&	Power of 2						\\ \specialrule{0em}{1pt}{1pt}
		$\chi_{\beta}$	&	Discrete Gaussian distribution with scalar $\beta$\\ \specialrule{0em}{1pt}{1pt}
		$re, se, te$	&	Random sampled values from $\chi_{\beta}$	\\ \specialrule{0em}{1pt}{1pt}
		$R_q$			&	Polynomial ring	\\ \specialrule{0em}{1pt}{1pt}
		$sk_{ncc}$	&	Private key of NCC		\\ \specialrule{0em}{1pt}{1pt}
		$pk_{j}, sk_{j}$	&	Public and private key pair of $L_j$		\\ \specialrule{0em}{1pt}{1pt}
		$pk_{i}, sk_{i}$	&	Public and private key pair of $u_i$		\\ \specialrule{0em}{1pt}{1pt}
		$pk_{tcs}, sk_{tcs}$	&	Public and private key pair of TCS	\\ \specialrule{0em}{1pt}{1pt}
		$Gen(\cdot), Rep(\cdot)$		&	Two functions of biometric fuzzy extractor\\ \specialrule{0em}{1pt}{1pt}
		$\sigma_i$		&	Biometric key of $u_i$			\\ \specialrule{0em}{1pt}{1pt}
		$v_i$			&	Auxiliary data of $\sigma_i$	\\ \specialrule{0em}{1pt}{1pt}
		$h(\cdot)$		&	Secure hash function $h:\{0,1\}^*\rightarrow \{0,1\}^l$ \\ \specialrule{0em}{1pt}{1pt}
		$H(\cdot)$		&	Secure hash function $H:\{0,1\}^*\rightarrow R_q$ \\
		\specialrule{0em}{1pt}{1pt}
		$t,T$			&	Timestamp						\\ \specialrule{0em}{1pt}{1pt}
		$\oplus$		&	XOR operation					\\ \specialrule{0em}{1pt}{1pt}
		$key$			&	Session key						\\ \specialrule{0em}{1pt}{1pt}
		\hline       
	\end{tabular}
	\label{Symbol list}
\end{table}

\subsection{System Initialization Phase}
NCC generates system public parameters and the master private key through the following steps.
\begin{itemize}
	\item [(1)]
	NCC first chooses an integer $n$ that is a power of 2 and then chooses the prime number $q$ such that $q \bmod n=1$.
	\item [(2)]
	NCC chooses the discrete Gaussian distribution $\chi_{\beta}$ with the standard deviation $\beta$ on polynomial ring $R_q$.
	\item [(3)]
	Next, NCC chooses two secure hash functions $h:\{0,1\}^*\rightarrow \{0,1\}^l$ and $H:\{0,1\}^*\rightarrow R_q$, and two functions of the biometric fuzzy extractor, namely $Gen(\cdot)$ and $Rep(\cdot)$.  
	\item [(4)]
	Then NCC randomly selects two elements $a$ and $sk_{ncc}$ from $\chi_{\beta}$. 
	\item [(5)]
	Finally, NCC uses $sk_{ncc}$ as the master private key and stores it confidentially, then publishes public parameters and functions $\{n, q, \chi_{\beta}, a, h(\cdot), H(\cdot), Gen(\cdot), Rep(\cdot)\}$ to the entire system.

\end{itemize} 

\subsection{Registration Phase}
The registration phase is divided into two categories by entity type, one is TCS and satellite node registration, the other is user registration. 
During the registration phase of TCS and satellite node, TCS and satellite node submit registration requests to NCC to obtain public-private key pairs issued by NCC. Similarly, 
in the user registration phase, the user submits a registration request to the TCS that provides a specific subscription services to obtain the authentication parameters required for access. It is worth mentioning that during the registration phase all messages are transmitted over secure channels. The detailed steps for the registration phase are described below. 

TCS registration. First, TCS randomly samples $sk_{tcs}$ and $se_{tcs}$ from $\chi_{\beta}$, and calculates $pk_{tcs}=a\cdot sk_{tcs}+2\cdot se_{tcs}$, where $pk_{tcs}$ and $sk_{tcs}$ are the master public-private key pairs of TCS. Then TCS sends its identity $ID_{tcs}$ and $pk_{tcs}$ to NCC. Finally, NCC calculates $p_{ncc}=h(sk_{ncc},T)$, where $T$ is the expiration timestamp, then publicly publishes $\{ID_{tcs},pk_{tcs},T\}$ to the entire system. TCS securely stores $\{sk_{tcs},p_{ncc}\}$.

Satellite node registration. The registration process is similar to TCS registration. First, $L_j$ randomly samples master private key $sk_{j}$ and $se_j$ from $\chi_{\beta}$, calculates master public key $pk_j=a\cdot sk_j+2\cdot se_{j}$, and sends $\{ID_j,pk_j\}$ to NCC. 
Then NCC publicly publishes $\{ID_j,pk_j, T\}$ to the entire system. Finally, $L_j$ securely stores $\{sk_{j}, p_{ncc}\}$.

When the validity of TCS and satellite nodes reach the expiration time $T$, the registration request needs to be resubmitted to NCC. 
\begin{figure}[ht] 
	\centering  
	\includegraphics[width=0.5\textwidth]{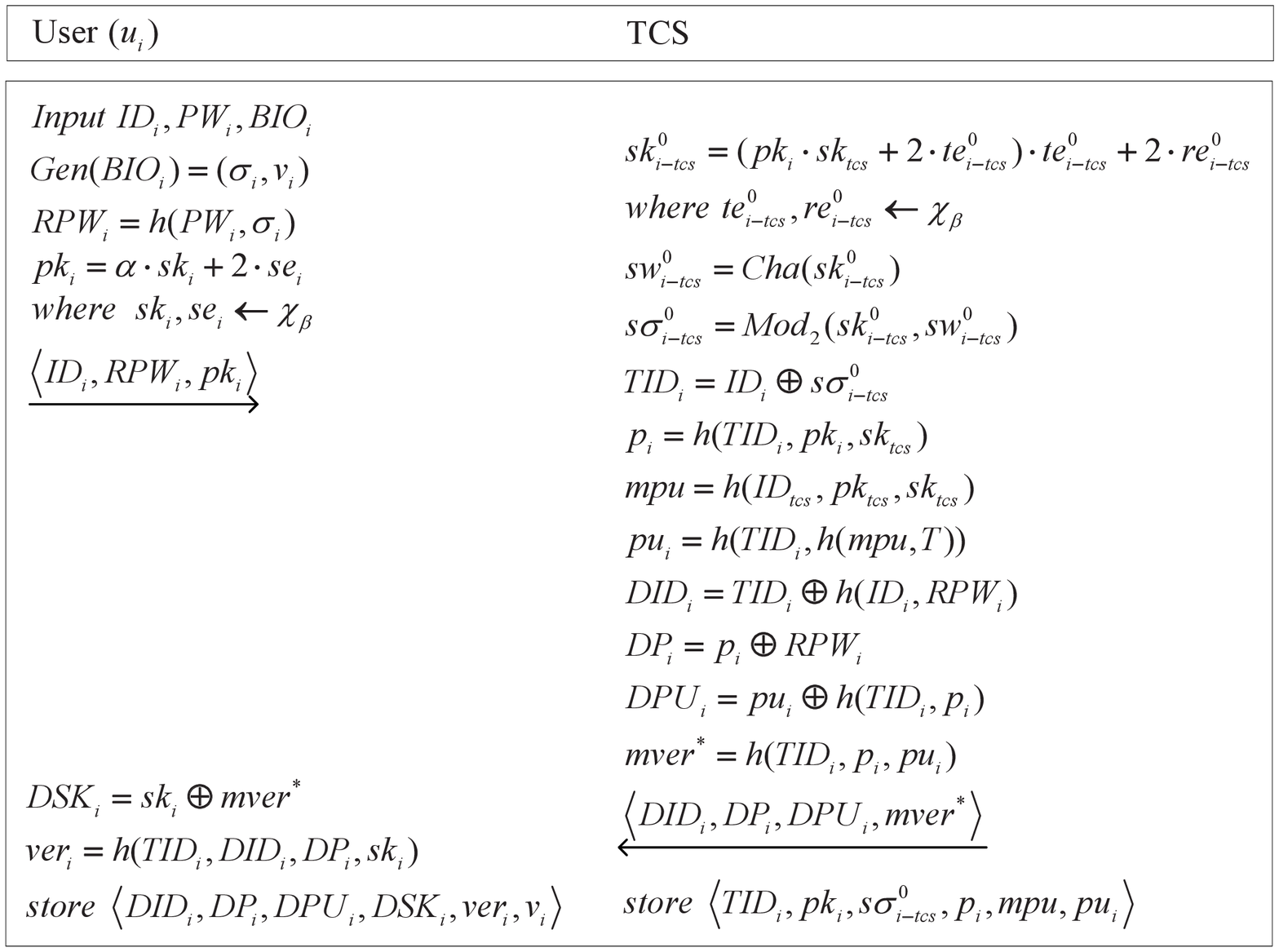}  
	\caption{User registration}  
	\label{userregistration}  
\end{figure}

User registration. As shown in Figure \ref{userregistration}, $u_i$ registers with TCS through the following steps.

\begin{itemize}
	\item [(1)]
	$u_i$ enters the identity $ID_i$, password $PW_i$, and biometric $BIO_i$, then executes the probabilistic generation function $Gen(BIO_i)$ to obtain the biometric key $\sigma_i$ and the public auxiliary data $v_i$. Next, $u_i$ calculates $RPW_i=h(PW_i,\sigma_i)$ and master public key $pk_i=a\cdot sk_i+2\cdot se_i$, where $sk_i$ and $se_i$ are randomly sampled from $\chi_{\beta}$ and $sk_i$ is the master private key. Finally, $u_i$ sends the registration request $\{ID_i,RPW_i,pk_i\}$ to TCS.
	\item [(2)]
	After receiving the registration request from $u_i$, TCS calculates $sk^{0}_{i-tcs}=(pk_i\cdot sk_{tcs} + 2\cdot te^{0}_{i-tcs})\cdot te^{0}_{i-tcs} + 2\cdot re^{0}_{i-tcs}$, where $te^{0}_{i-tcs}$ and $re^{0}_{i-tcs}$ are randomly sampled from $\chi_{\beta}$. Next, TCS calculates $sw^{0}_{i-tcs}=Cha(sk^{0}_{i-tcs})$, $s\sigma^{0}_{i-tcs}=Mod_2(sk^{0}_{i-tcs},sw^{0}_{i-tcs})$, $u_i$'s temporary identity $TID_i=ID_i\oplus s\sigma^{0}_{i-tcs}$, $p_i=h(TID_i,pk_i,sk_{tcs})$, $mpu=h(ID_{tcs},pk_{tcs},sk_{tcs})$, $pu_i=h(TID_i,h(mpu,T))$, $DID_i=TID_i\oplus h(ID_i,RPW_i)$, $DP_i=p_i\oplus RPW_i$, $DPU_i=pu_i\oplus h(TID_i,p_i)$, $mver^{*}=h(TID_i,p_i,pu_i)$. Finally, TCS returns the registration response $\{DID_i,DP_i,DPU_i,mver^{*}\}$ to $u_i$ and securely stores $\{TID_i,pk_i,s\sigma^{0}_{i-tcs},p_i,mpu,pu_i\}$.
	\item [(3)]
	$u_i$ calculates $DSK_i=sk_i\oplus mver^{*}$, $ver_i=h(TID_i,DID_i,DP_i,sk_i)$ after receiving the registration response, where $ver_i$  is used to verify the correctness of the identity $ID_i$, password $PW_i$ and biometric $BIO_i$ entered by the user when logging in, and to check that the authentication parameters stored in the device have not been modified. Finally, $u_i$ stores $\{DID_i,DP_i,DPU_i,DSK_i,ver_i,v_i\}$ in the device.

\end{itemize}

\subsection{Pre-Negotiation Phase}

\begin{figure}[ht] 
	\centering  
	\includegraphics[width=0.5\textwidth]{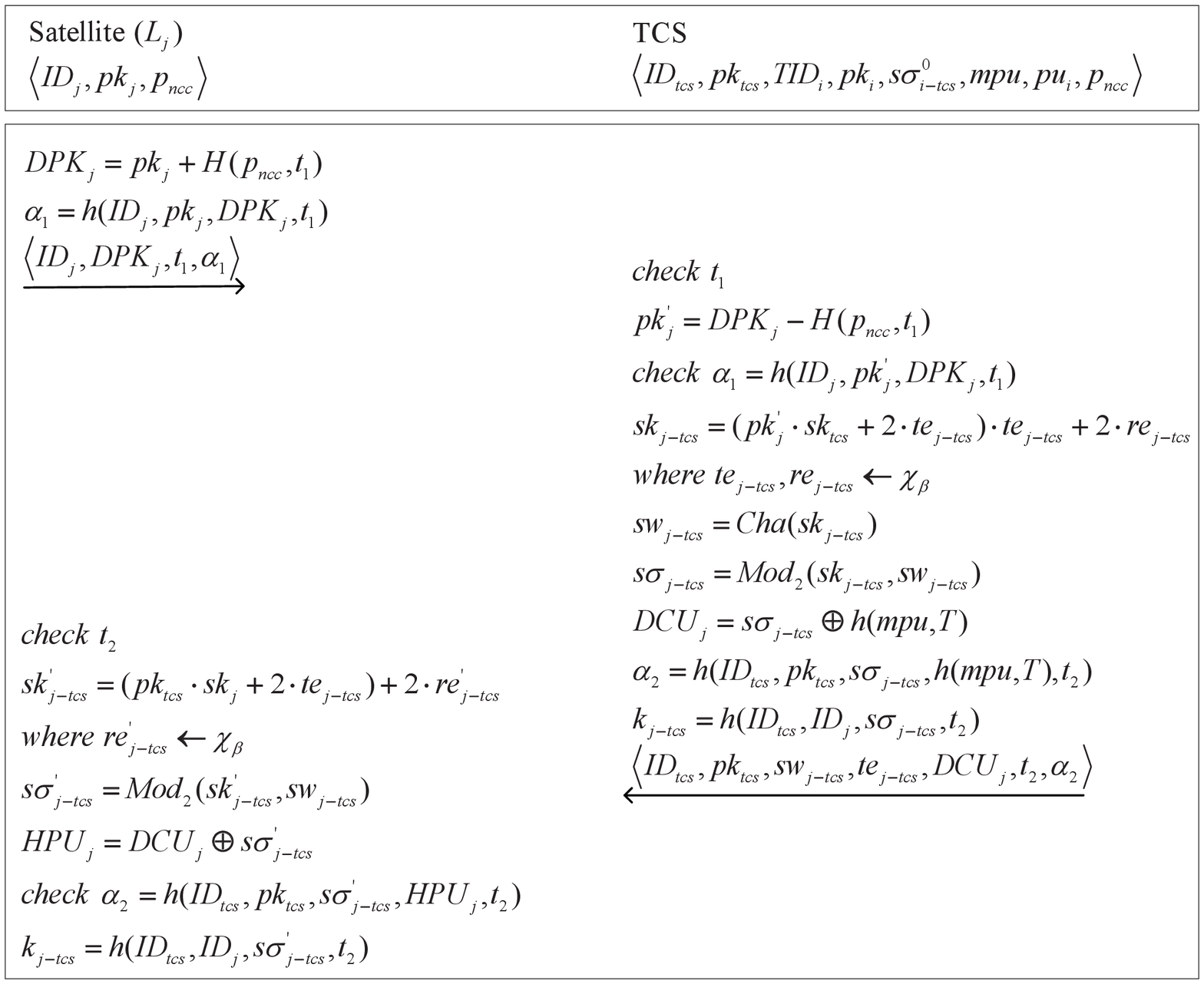}  
	\caption{Pre-Negotiation phase}  
	\label{Pre-Negotiation}  
\end{figure}

Figure \ref{Pre-Negotiation} shows the pre-negotiation phase between the satellite and TCS. 
The satellite sends a pre-negotiation request to negotiate authentication parameters for future user access and construct a secure channel with TCS.
Pre-negotiation reduces the computational and communication overhead of user access authentication phase to meet the quality of service requirements of time-sensitive users. The detailed steps of pre-negotiation phase are as follows.

\begin{figure*}[ht] 
	\centering  
	\includegraphics[width=\textwidth]{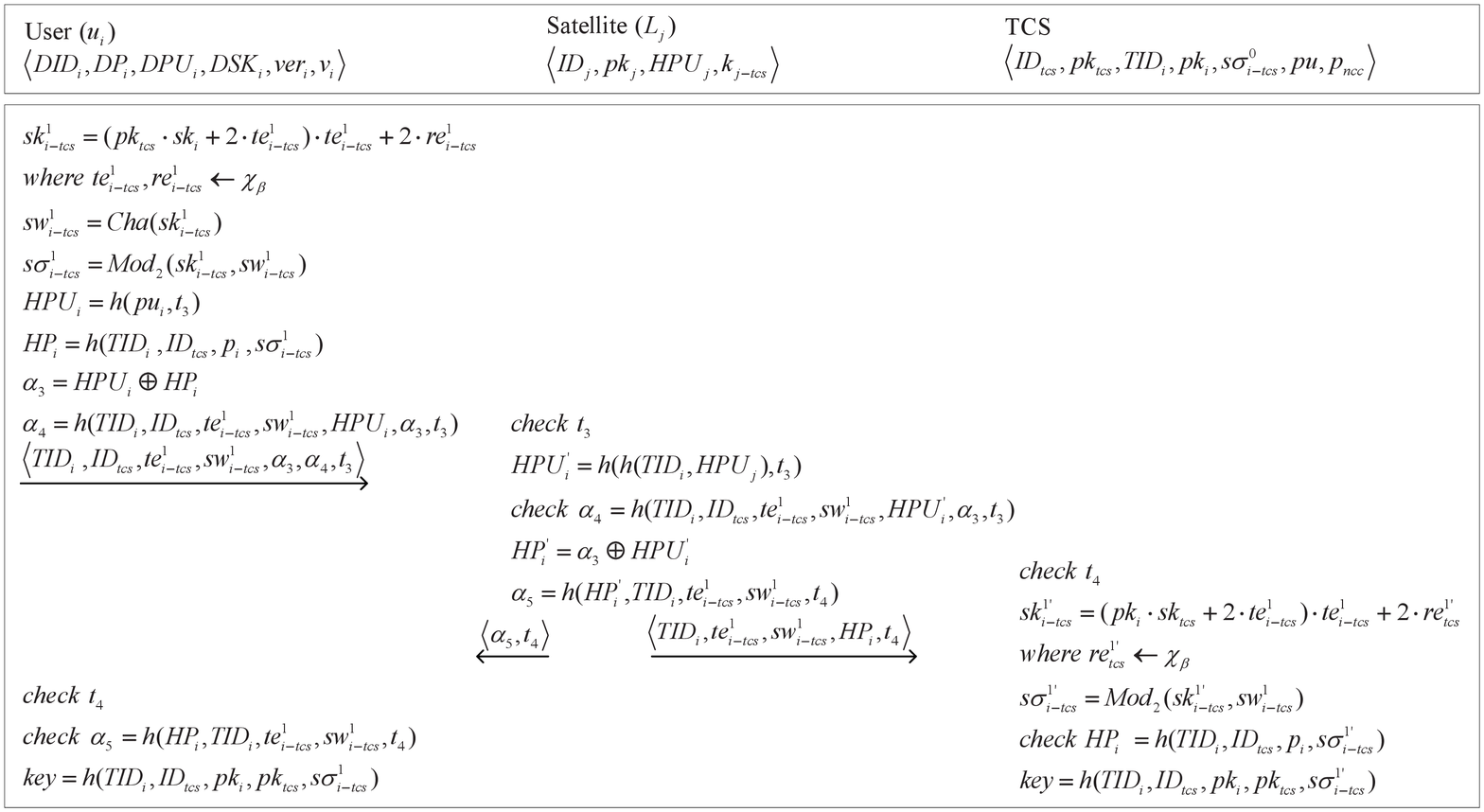}  
	\caption{Authentication process}  
	\label{Authentication Phase}
\end{figure*}

\begin{itemize}
	\item[(1)] 
	$L_j$ calculates $DPK_j=pk_j+H(p_{ncc},t_1)$, $a_1=h(ID_j,pk_j,DPK_j,t_1)$, where $t_1$ is the current timestamp, and then sends the pre-negotiation message $\{ID_j,DPK_j,t_1,a_1\}$ to TCS.
	
	\item[(2)]
	After receiving the pre-negotiation message $\{ID_j,DPK_j,t_1,a_1\}$, TCS first checks the freshness of $t_1$ by verifying whether the timestamp is within the allowable transmission delay threshold. Next, TCS calculates $pk^{'}_j=DPK_j-H(p_{ncc},t_1)$ and checks whether $a_1=h(ID_j,pk^{'}_j,DPK_j,t_1)$ holds. If the check passes, the pre-negotiation message sent by $L_j$ has not been tampered with and is indeed sent by the legitimate satellite node, otherwise terminate the pre-negotiation. Then TCS randomly samples $te_{j-tcs}$ and $re_{j-tcs}$ from $\chi_{\beta}$ and calculates $sk_{j-tcs}=(pk^{'}_j\cdot sk_{tcs} + 2\cdot te_{j-tcs})\cdot te_{j-tcs}+2\cdot re_{j-tcs}$, $sw_{j-tcs}=Cha(sk_{j-tcs})$, $s\sigma_{j-tcs}=Mod_2(sk_{j-tcs},sw_{j-tcs})$, $DCU_j=s\sigma_{j-tcs}\oplus h(mpu,T)$, $a_2=h(ID_{tcs},pk_{tcs},s\sigma_{j-tcs},h(mpu,T),t_2)$ and the session key between $L_j$ and TCS is $k_{j-tcs}=h(ID_{tcs},ID_j,s\sigma_{j-tcs},t_2)$, where $t_2$ is the timestamp, $k_{j-tcs}$ is used to construct the secure channel between $L_j$ and TCS to protect the security of future mutual communication during access authentication phase and communication. Finally, TCS sends the pre-negotiation response $\{ID_{tcs},pk_{tcs},sw_{j-tcs},te_{j-tcs},DCU_j,t_2,a_2\}$ to $L_j$.
	
	\item [(3)]
	After receiving the message $\{ID_{tcs},pk_{tcs},sw_{j-tcs},te_{j-tcs},DCU_j,t_2,a_2\}$, $L_j$ first checks the timestamp $t_2$, then randomly samples $re^{'}_{j-tcs}$ from $\chi_{\beta}$ and calculates $sk^{'}_{j-tcs}=(pk_{tcs}\cdot sk_{j} + 2\cdot te_{j-tcs}) + 2\cdot re^{'}_{j-tcs}$,  $s\sigma^{'}_{j-tcs}=Mod_2(sk^{'}_{j-tcs},sw_{j-tcs})$ and $HPU_j=DCU_j\oplus s\sigma^{'}_{j-tcs}$. Next, $L_j$ checks whether $a_2=h(ID_{tcs},pk_{tcs},s\sigma^{'}_{j-tcs},HPU_j,t_2)$ holds. If the equation holds, the pre-negotiation response message is not modified and $L_j$ confirms that the sender is the legitimate node. Finally, $L_j$ calculates $k_{j-tcs}=h(ID_{tcs},ID_{j},s\sigma^{'}_{j-tcs},t_2)$, which is used as the session key to construct a secure communication channel with TCS.
\end{itemize}

\subsection{Login and Authentication Phase}

Before accessing SIN and submitting the access request, the user needs to complete the three-factor login verification, that is, 
verify the parameters stored in the smart device and the identity, password and biometric entered by the user.
Login verification can prevent the adversary from using the user's lost device to send a large number of access requests and consume the computing and communication resources of the satellite and TCS in SIN.
The detailed steps of login and authentication phase are as follows, and the authentication phase is shown in Figure \ref{Authentication Phase}.

\begin{itemize}
	\item [(1)] 
	$u_i$ enters the identity $ID^{*}_i$, password $PW^{*}_{i}$ and scans the biometric $BIO^{*}_{i}$, and then obtains the biometric key $\sigma^{*}_i$ through the reproduction function of biometric fuzzy extractor and the public auxiliary data $v_i$. Next, $u_i$ calculates $RPW^{'}_i=h(PW^{*}_{i},\sigma^{*}_i)$, $TID^{'}_i=DID_i\oplus h(ID^{*}_i,RPW^{'}_i)$, $p^{'}_i=DP_i\oplus RPW^{'}_i$, $pu^{'}_i=DPU_i\oplus h(TID^{'}_i,p^{'}_i)$, $sk^{'}_i=DSK_i\oplus h(TID^{'}_i,p^{'}_i,pu^{'}_i)$, and checks whether $ver_i=h(TID^{'}_i,DID_i,DP_i,sk^{'}_i)$ holds. If the check passes, it means that $ID^{*}_i$, $PW^{*}_{i}$ and $BIO^{*}_{i}$ entered by $u_i$ are correct and the parameters $\{DID_i,DP_i,DPU_i,DSK_i,ver_i,v_i\}$ stored in the device have not been tampered with, otherwise the login is terminated. After that, $u_i$ randomly samples $te^{1}_{i-tcs}$ and $re^{1}_{i-tcs}$ from $\chi_{\beta}$ and calculates $sk^{1}_{i-tcs}=(pk_{tcs}\cdot sk_i+2\cdot te^{1}_{i-tcs})\cdot te^{1}_{i-tcs}+2\cdot re^{1}_{i-tcs}$, $sw^{1}_{i-tcs}=Cha(sk^{1}_{i-tcs})$, $s\sigma^{1}_{i-tcs}=Mod_2(sk^{1}_{i-tcs},sw^{1}_{i-tcs})$, $HPU_i=h(pu_i,t_3)$, $HP_i=h(TID_i,ID_{tcs}, p_i, s\sigma^{1}_{i-tcs})$, $a_3=HPU_i\oplus HP_i$, $a_4=h(TID_i,ID_{tcs},te^{1}_{i-tcs},sw^{1}_{i-tcs},HPU_i,a_3,t_3)$, where $t_3$ is the timestamp. Finally, $u_i$ sends the access request $\{TID_i,ID_{tcs},te^{1}_{i-tcs},sw^{1}_{i-tcs},a_3,a_4,t_3\}$ to $L_j$. 
	
	\item [(2)] 
	After receiving the access request from $u_i$, $L_j$ first checks the timestamp $t_2$ of the message, calculates $HPU^{'}_i=h(h(TID_i,HPU_j),t_3)$ and verifies whether $a_4=h(TID_i,ID_{tcs},te^{1}_{i-tcs},sw^{1}_{i-tcs},HPU^{'}_i,a_3,t_3)$ is established. If the verification is passed, the access request received through the non-secure channel has not been maliciously modified and comes from the legitimate node $u_i$. Next, $L_j$ calculates $HP^{'}_i=a_3\oplus HPU^{'}_i$, $a_5=h(HP^{'}_i,TID_i,te^{1}_{i-tcs},sw^{1}_{i-tcs},t_4)$, where $t_4$ is the current timestamp. Finally, $L_j$ simultaneously sends $\{a_5,t_4\}$ and $\{TID_i,te^{1}_{i-tcs},sw^{1}_{i-tcs},HP_i,t_4\}$ to $u_i$ and TCS, respectively.
	
	\item [(3)]
	After receiving the message $\{a_5,t_4\}$ from $L_j$, $u_i$ checks the timestamp $t_4$ and whether $a_5=h(HP_i,TID_i,te^{1}_{i-tcs},sw^{1}_{i-tcs},t_4)$ is established. If the check passes, $u_i$ calculates the session key $key=h(TID_i,ID_{tcs},pk_i,pk_{tcs},s\sigma^{1}_{i-tcs})$.
	
	\item[(4)] 
	TCS receives $\{TID_i,te^{1}_{i-tcs},sw^{1}_{i-tcs},HP_i,t_4\}$ from $L_j$, first checks the timestamp $t_4$ and randomly samples $re^{1'}_{i-tcs}$ from $\chi_{\beta}$, then calculates $sk^{1'}_{i-tcs}=(pk_i\cdot sk_{tcs}+2\cdot te^{1}_{i-tcs}) \cdot te^{1}_{i-tcs} +2\cdot re^{1'}_{tcs}$, $s\sigma^{1'}_{i-tcs}=Mod_2(sk^{1'}_{i-tcs},sw^{1'}_{i-tcs})$. Next, TCS checks whether $HP_i=h(TID_i,ID_{tcs},p_i,s\sigma^{1'}_{i-tcs})$ holds. If the equation holds, TCS confirms that $u_i$ is the legitimate user and then calculates the session key $key=h(TID_i,ID_{tcs},pk_i,pk_{tcs},s\sigma^{1'}_{i-tcs})$.
		
\end{itemize}

So far, the satellite and TCS authenticate the legitimacy of the access user. Similarly, the user confirms the legitimacy of the satellite node. Furthermore, the session key $key$ will protect the future communication between the TCS and the user.

\subsection{Handover Phase}
\begin{figure}[ht] 
	\centering  
	\includegraphics[width=0.5\textwidth]{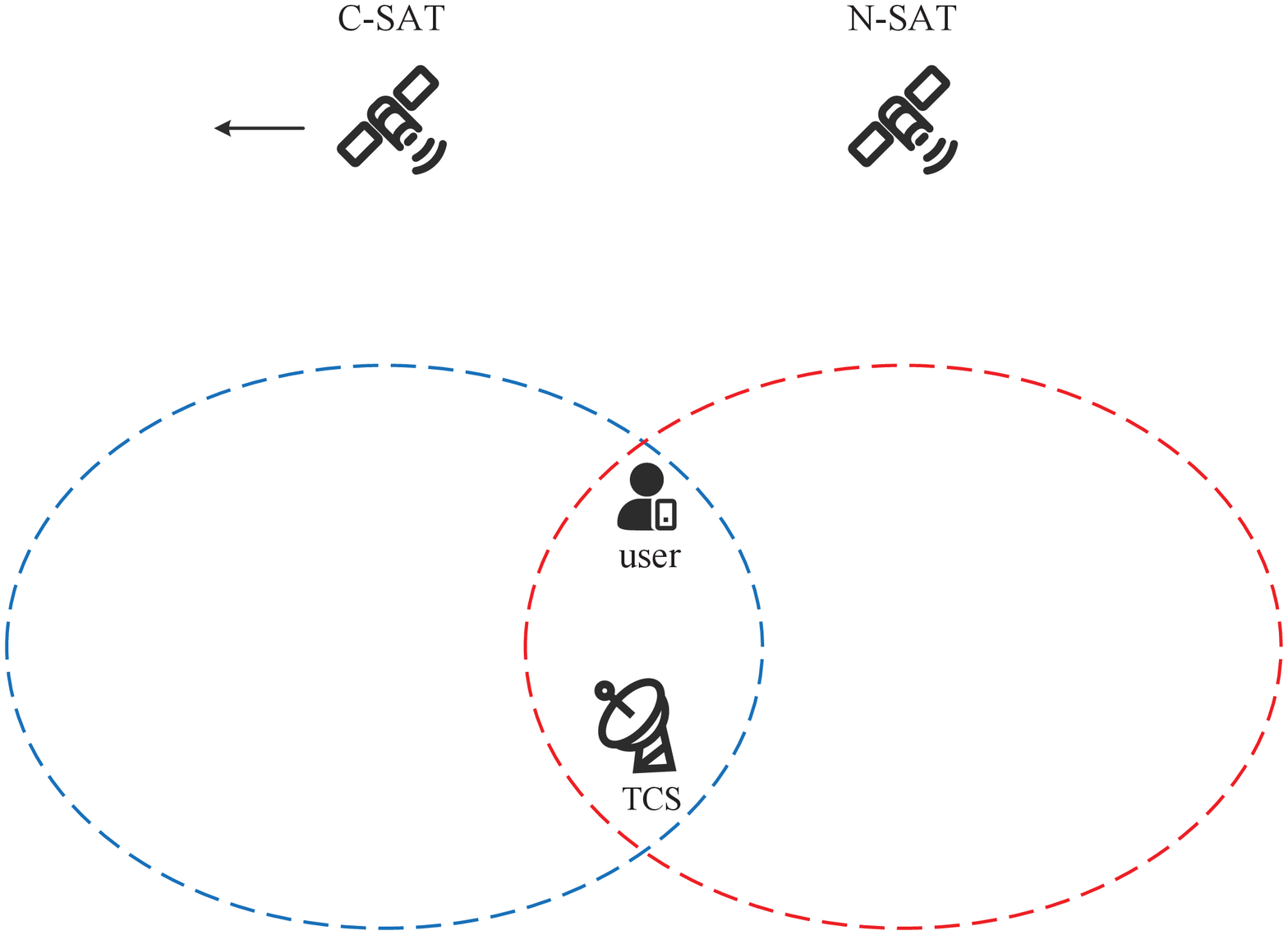}  
	\caption{Handover scenario}  
	\label{handover}
\end{figure}

As shown in Figure \ref{handover}, the blue circle is the current satellite (C-SAT) signal coverage, and the red circle is the next satellite (N-SAT) signal coverage. Due to the mobility of C-SAT, the signal strength received by the user and TCS gradually decreases. At this time, the user and TCS need to make signal handover in the overlapping areas of C-SAT and N-SAT satellite signal coverage. In the handover process, only relay satellite replacement is involved, which means that the user and TCS do not need to renegotiate the new session key but only need to mutually authenticate with N-SAT separately. 
Although Xue et al. \cite{Xuehandover} proposed a handover verification scheme based on the whitelist mechanism for such handover scenario, there are many shortcomings as follows. 
First, the user does not confirm the validity of the N-SAT relay satellite before handing over the signal. 
Secondly, the user's handover request can be easily captured and replayed by an adversary to obtain the N-SAT signal relay service.
Finally, since the relay satellite typically serve thousands of users, using the whitelist mechanism can greatly consume the valuable storage capacity of the satellite. 
Therefore, to overcome the above shortcomings, we propose a handover scheme that supports mutual authentication between the user and the N-SAT, resists replay attacks and does not consume the storage overhead of the N-SAT satellite.
The detailed steps of the handover phase are as follows.


\begin{itemize}
	\item[(1)] 
	Pre-negotiation. Similar to the pre-negotiation phase in the third subsection of this section, TCS and N-SAT mutually confirm the legal identity of each other through pre-negotiation, and TCS transmits to N-SAT the authentication parameter $HPU_{N-SAT}$ that enable N-SAT and the user to mutually authenticate. 
	
	\item[(2)] 
	$u_i$ sends the handover request. After $u_i$ enters the correct identity $ID_i$, password $PW_i$ and biometric $BIO_i$, then randomly samples $HO_i$ from $\chi_{\beta}$ and calculates $HPU_i=h(pu_i,t_5)$, $HPO_i=HPU_i\oplus h(HO_i,t_5)$, $a_6=h(TID_i,ID_{N-SAT},HPO_i,HPU_i,t_5)$, where $t_5$ is the timestamp. Finally, $u_i$ sends the handover request $\{TID_i,ID_{N-SAT},HPO_i,a_6,t_5\}$ via C-SAT to N-SAT. 
	
	\item[(3)]
	N-SAT verification. N-SAT first checks the freshness of the timestamp $t_5$, then calculates $HPU^{'}_i=h(h(TID_i,HPU_{N-SAT}),t_5)$ and checks whether $a_6=h(TID_i,ID_{N-SAT},HPO_i,HPU^{'}_i,t_5)$ holds. If the check passes, the handover request is not tampered with and $u_i$ is a legitimate user. Next, N-SAT calculates 
	$a_7=h(TID_i,ID_{N-SAT},HPO_i\oplus HPU^{'}_i,t_6)$ and sends the request response $\{a_7,t_6\}$ to $u_i$.
	
	\item[(4)] 
	$u_i$ verification. $u_i$ checks the timestamp $t_6$ and whether $a_7=h(TID_i,ID_{N-SAT},h(HO_i,t_5),t_6)$ holds. If the check passes, $u_i$ confirms that N-SAT is a valid relay node and the future mutual communication between the user and TCS will be relayed by N-SAT.
\end{itemize}

In the third and fourth steps of the handover process, the N-SAT satellite and the user mutually authenticate each other for legitimacy. 
And since the handover messages include timestamps and one-way hash check values, the adversary cannot obtain relay services from the satellite through replay attacks.
Furthermore, the N-SAT satellite only needs three hashing operations and does not need to store the whitelist but only stores the authentication parameter $HPU_{N-SAT}$ negotiated with TCS in the pre-negotiation phase, which greatly reduces the computing and storage load of the resource-constrained relay satellite.

\subsection{Password and Biometric Update Phase}
For some security purposes, the user can update his or her password and biometric without interacting with TCS, which can greatly reduce the communication and computation overhead of TCS. The detailed steps of the password and biometric update phase are as follows. 
\begin{itemize}
	\item[(1)] 
	$u_i$ enters the identity $ID^{*}_i$, old password $PW^{old}_i$, and biometric $BIO^{old}_i$ then obtains the biometric key $\sigma^{old}_i$ through $Rep$ function.
	\item[(2)] 
	The device calculates $RPW^{'}_i=h(PW^{old}_{i},\sigma^{old}_i)$, $TID^{'}_i=DID_i\oplus h(ID^{*}_i,RPW^{'}_i)$, $p^{'}_i=DP_i\oplus RPW^{'}_i$, $pu^{'}=DPU_i\oplus h(TID^{'}_i,p^{'}_i)$, $sk^{'}_i=DSK_i\oplus h(TID^{'}_i,p^{'}_i,pu^{'}_i)$, and checks whether $ver_i=h(TID^{'}_i,DID_i,DP_i,sk^{'}_i)$ holds. If the check passes, continue to the next step, otherwise terminate the update. 
	\item[(3)]
	$u_i$ enters new password $PW^{new}_i$ and biometric $BIO^{new}_i$, then obtains the new biometric key $\sigma^{new}_i$ and auxiliary data $v^{new}_i$ through $Gen$ function. 
	\item[(4)]
	The device calculates $RPW^{new}_i=h(PW^{new}_{i},\sigma^{new}_i)$, $DID^{new}_i=TID_i\oplus h(ID_i, RPW^{new}_i)$, $DP^{new}_i=p_i\oplus RPW^{new}_i$, $ver^{new}_i=h(TID_i,DID^{new}_i,DP^{new}_i,sk_i)$ and finally deletes \{$DID^{old}_i$, $DP^{old}_i$, $v^{old}_i$, $ver_i$\} and stores \{$DID^{new}_i$, $DP^{new}_i$, $v^{new}_i$, $ver^{new}_i$\} in the device. After that, $u_i$ can log in with the new password $PW^{new}_{i}$ and new biometric $BIO^{new}_{i}$.
	
\end{itemize}

\section{Security Analysis} \label{SecurityAnalysis}

In this section, we first formally prove the security of our proposed PSAA protocol based on the widely accepted random oracle model (ROM) \cite{ROMBellare, ROMChatter} and the automated validation of internet security protocols and applications (AVISPA) tool \cite{AVISPA}, in which ROM proves the semantic security of the negotiated session key, and AVISPA proves that the PSAA protocol can resist eavesdropping attacks, replay attacks and man-in-the-middle attacks. Then the informal security analysis method discusses that PSAA protocol can meet the security requirements mentioned in Section 4.

\subsection{Formal Security Analysis Based on ROM}

According to protocol we proposed, there are a total of three instances in the authentication phase, namely $u_i$, $L_j$ and TCS. Based on the threat model, the adversary $\mathcal{A}$ is defined as a Turing machine that attempts to break the PSAA protocol in probabilistic polynomial time. The following definitions need to be made before ROM-based proof. 

Participants. The symbols $\prod^{x}_{u_i}$, $\prod^{y}_{L_j}$ and $\prod^{z}_{TCS}$ represent the $x$-th, $y$-th, and $z$-th instances of $u_i$, $L_j$, and TCS, respectively. Each instance is an oracle, that is, after receiving the correct authentication message, it will respond honestly.


Accepted state. After instance $\prod^{x}$ receives the last correct expected message according to the steps of authentication phase, the instance is in the accepted state. The session identification $sid$ represents all the messages received and sent in sequence in the current session of instance $\prod^{x}$. 

Freshness. If the current session key of instance $\prod^{x}$ or its partner $\prod^{y}$ is not obtained by adversary $\mathcal{A}$, then instance $\prod^{x}$ or $\prod^{y}$ is said to be fresh.

Partnering. Instances $\prod^{x}$ and $\prod^{y}$ become partners when they meet the following three conditions simultaneously. 
\begin{itemize}
	\item $\prod^{x}$ and $\prod^{y}$ are all in the accepted state.
	\item $\prod^{x}$ and $\prod^{y}$ have the same session identification $sid$ and mutually authenticate each other. 
	\item $\prod^{x}$ and $\prod^{y}$ are each other's partners \cite{Srinivaschaotic}.
\end{itemize}

Adversary. As described in the threat model, $\mathcal{A}$ can not only intercept all messages transmitted in the public channel and modify, forge and replay them, but also obtain information stored in the user's smart device through power analysis. We define the capabilities of $\mathcal{A}$ as the following query oracles. 
\begin{itemize}
	\item
	Execute($\prod^{x}_{u_i}$, $\prod^{y}_{L_j}$, $\prod^{z}_{TCS}$). $\mathcal{A}$ can obtain all authentication messages transmitted by instances $\prod^{x}_{u_i}$, $\prod^{y}_{L_j}$ and $\prod^{z}_{TCS}$ in the public channel during the authentication process through this query. The query oracle can be regarded as an eavesdropping attack.
	\item
	Send($\prod^{x}$, M). Through this query, $\mathcal{A}$ can send a message M to instance $\prod^{x}$, where M can be a message obtained by eavesdropping, or a modified or forged message. 
	\item
	Reveal($\prod^{x}$). $\mathcal{A}$ can obtain the current session key of instance $\prod^{x}$ and its partner $\prod^{y}$ through the Reveal query. 
	\item 
	$h$(M). $h$ oracle maintains a hash list $L_h: \{M_i, \zeta_i\}$, where $\zeta_i$ is the hash value of message $M_i$. When $\mathcal{A}$ queries $h$ oracle for the hash value of the message M, if M is in $L_h$, returns $\zeta$, otherwise $\zeta$ is randomly selected from $\{0,1\}^l$, then $\zeta$ is stored in $L_h$ and returned to  $\mathcal{A}$.
	\item
	CorruptSmartDevice($\prod^{x}$). $\mathcal{A}$ can obtain the information stored in instance $\prod^{x}$ through this query. This query can be regarded as a power analysis attack or a side channel attack on the user's smart device \cite{ROMChatter,ChallaECC-based}.
	\item 
	Test($\prod^{x}$). Test query is to check whether the current session key $key$ of $\prod^{x}$ meets semantic security under the random oracle model. When $\mathcal{A}$ queries the test oracle, $\prod^{x}$ first chooses an unbiased coin $C\in\{0,1\}$ and  flip it. If $C=1$, $\prod^{x}$ returns the current session key $key$ to $\mathcal{A}$, and if $C=0$, $\prod^{x}$ returns to $\mathcal{A}$ a random string $str$ of the same length as $key$, otherwise returns null.
\end{itemize}

Semantic security of session key. In ROM, if $\mathcal{A}$ cannot distinguish the true session key from the session key and a random string of the same length as the session key, it means that  $\mathcal{A}$ cannot compromise the semantic security of the session key. When testing the semantic security of the session key,  $\mathcal{A}$ can query Test($\prod^{x}$) multiple times. Each time $\prod^{x}$ re-selects an unbiased coin $C$ and flip it, then $\mathcal{A}$ guesses the true value of the coin is $C^{'}$ through the above query. If $C=C^{'}$, $\mathcal{A}$ wins the game. We define the probability of $\mathcal{A}$ to win the game as $Pr(Succ)$, and the advantage of breaking the protocol PSAA is $Adv^{\mathcal{A}}_{PSAA}=\vert2Pr(Succ)-1\vert$. When $Adv^{\mathcal{A}}_{PSAA}<\epsilon$, the protocol PSAA is secure and $\mathcal{A}$ cannot compromise the semantic security of the session key, where $\epsilon$ is a negligible value.

\textbf{Theorem 1.} Suppose that $\mathcal{A}$ is a Turing machine in probability polynomial time attempting to break the protocol PSAA and obtain the session key. $q_h$, $q_s$ and $q_e$ respectively represent the maximum number of times $\mathcal{A}$ can query $h$, Send, Execute. Besides, $\vert D \vert$ is the size of the user password set, $l_1$ is the length of the biometric key, $\vert \chi_{\beta}\vert$ is the distribution range of Gaussian distribution $\chi_{\beta}$, $\vert h \vert$ is the output range of the hash function $h$, $Adv^{\mathcal{A}}_{RLWE}$ is the advantage that $\mathcal{A}$ can solve the RLWE assumption in polynomial time. Then the advantage of $\mathcal{A}$ who breaks the protocol PSAA and obtains the session key is defined as $Adv^{\mathcal{A}}_{PSAA}$, and when $Adv^{\mathcal{A}}_{PSAA}  \le \frac{q^{2}_h}{\vert h\vert} + \frac{(q_s + q_e)^2}{\vert \chi_{\beta}\vert} + \frac{q_s}{2^{l_1- 1}\vert D \vert } + 2Adv^{\mathcal{A}}_{RLWE}$ holds, it can be considered that our proposed protocol PSAA is secure.

$Proof$: In order to evaluate the advantage of $\mathcal{A}$ in breaking the semantic security of the key, we execute a series of games, namely $Game_0$, $Game_1$, $Game_2$, $Game_3$ and $Game_4$. We define $Succ_i$ to represent the event in which $\mathcal{A}$ wins $Game_i$, and $Pr(Succ_i)$ to represent the probability of this event. The proof is similar to \cite{Ostad-Sharif, LBA_PAKE} and the details are as follows. 

$Game_0$. In this game, an unbiased coin $C$ is randomly selected and flipped. $\mathcal{A}$ implements the real attack on the protocol PSAA to guess the true value of the coin, then we get the following formula: 
	\begin{equation} 
		Adv^{\mathcal{A}}_{PSAA} =\vert 2Pr[Succ_0] - 1 \vert \label{equ1}
	\end{equation}

$Game_1$. Compared with $Game_0$, $\mathcal{A}$ performs Execute($\prod^{x}_{u_i}$, $\prod^{y}_{L_j}$, $\prod^{z}_{TCS}$) query to obtain the authentication message transmitted between the three entities in the non-secure channel. Through this query, $\mathcal{A}$ can only get $\{TID_i,ID_{tcs},te^{1}_{i-tcs},sw^{1}_{i-tcs},a_3,a_4,t_3\}$ and $\{a_5,t_4\}$. Since $key=h(TID_i,ID_{tcs},pk_i,pk_{tcs},s\sigma^{1}_{i-tcs})$ where $s\sigma^{1}_{i-tcs}=Mod_2(sk^{1}_{i-tcs},sw^{1}_{i-tcs})$, $\mathcal{A}$ cannot obtain $sk^{1}_{i-tcs}$ that can directly calculate $s\sigma^{1}_{i-tcs}$ from the intercepted message. Therefore, compared with $Game_0$, the advantage of breaking the protocol PSAA has not improved. Then we can get the following formula:
	\begin{equation} 
		Pr[Succ_1] = Pr[Succ_0] 				\label{equ2}
	\end{equation}
	
$Game_2$. In $Game_2$, $\mathcal{A}$ uses Send and $h$ queries to forge or modify authentication messages, and finally implements  active attacks such as forging or replaying messages. When the hash value of the message forged by $\mathcal{A}$ is the same as the authentication message, the secret value can be obtained through hash collision. According to the birthday paradox, the maximum hash collision probability of $h$ is $\frac{q^{2}_h}{2\vert h\vert}$. Furthermore, since $sk^{1}_{i-tcs}=(pk_{tcs}\cdot sk_i+2\cdot te^{1}_{i-tcs})\cdot te^{1}_{i-tcs}+2\cdot re^{1}_{i-tcs}$ where $te^{1}_{i-tcs}$ and $re^{1}_{i-tcs}$ are random sampling values in $\chi_{\beta}$, the maximum probability of a collision with a random value is $\frac{(q_s + q_e)^2}{2\vert \chi_{\beta}\vert}$. Then we can get the following formula: 
	\begin{equation} 
		\vert Pr[Succ_2] - Pr[Succ_1] \vert \le \frac{q^{2}_h}{2\vert h\vert} + \frac{(q_s + q_e)^2}{2\vert \chi_{\beta}\vert}   \label{equ3}
	\end{equation}
	
$Game_3$. Different from the previous game in that $\mathcal{A}$ can use CorruptSmartDevice  query to obtain the information $\{DID_i,DP_i,DPU_i,DSK_i,ver_i,v_i\}$ stored in the user's smart device and try to derive the user's password $PW_i$ and the biometric key $\sigma_i$ from the stored information. Once $\mathcal{A}$ has the correct user password and biometric key, he/she can pass the check of $ver_i$ and obtains the master private key $sk_i$. The probability of $\mathcal{A}$ guessing the password and the biometric key is $\frac{1}{\vert D \vert}$ and $\frac{1}{2^{l1}}$, respectively. In addition, $\mathcal{A}$ can only execute up to $q_s$ queries, and only when the password and the biometric key are correct at the same time can pass the verification, then we can get the following formula:
	\begin{equation} 
		\vert Pr[Succ_3] - Pr[Succ_2] \vert \le \frac{q_s}{2^{l_1}\vert D \vert }      \label{equ4}
	\end{equation}
	
$Game_4$. In this game, $\mathcal{A}$ attempts to obtain the session key $key=h(TID_i,ID_{tcs},pk_i,pk_{tcs},s\sigma^{1}_{i-tcs})$ by intercepting $\{te^{1}_{i-tcs},sw^{1}_{i-tcs},pk_i,pk_{tcs}\}$ in the public channel, where $s\sigma^{1}_{i-tcs}=Mod_2(sk^{1}_{i-tcs},sw^{1}_{i-tcs})$, $sk^{1}_{i-tcs}=(pk_{tcs}\cdot sk_i+2\cdot te^{1}_{i-tcs})\cdot te^{1}_{i-tcs}+2\cdot re^{1}_{i-tcs}$. $\mathcal{A}$ calculates $sk_i$, $sk_{tcs}$ and $sk^{1}_{i-tcs}$ from the public parameters $pk_i$ and $pk_{tcs}$, which can be reduced to solve the RLWE assumption. Then we can get the following formula: 
	\begin{equation} 
		\vert Pr[Succ_4] - Pr[Succ_3] \vert \le Adv^{\mathcal{A}}_{RLWE}       \label{equ5}
	\end{equation}

Finally, $\mathcal{A}$ has no other advantage to break the semantic security of the session key without the above method, so we can get the following formula:
\begin{equation} 
	Pr[Succ_4] = \frac{1}{2}												\label{equ6}
\end{equation}

According to formula \ref{equ1} and formula \ref{equ6}, we can get the following formula:
\begin{equation} 
	\begin{aligned} 
	Adv^{\mathcal{A}}_{PSAA} 	&= 2\vert Pr[Succ_0] -	 \frac{1}{2} \vert \\
							 	&= 2  \vert  Pr[Succ_0] -Pr[Succ_4] \vert 			\label{equ7}
	\end{aligned} 
\end{equation}

Next, according to the triangle inequality and formulas \ref{equ2}, \ref{equ3}, \ref{equ4}, \ref{equ5}, we can get the following formula:

\begin{equation}
	\begin{aligned} 
		Adv^{\mathcal{A}}_{PSAA} & \le \frac{q^{2}_h}{\vert h\vert} + \frac{(q_s + q_e)^2}{\vert \chi_{\beta}\vert} + \frac{q_s}{2^{l_1- 1}\vert D \vert } + 2Adv^{\mathcal{A}}_{RLWE}
	\label{equ8}
	\end{aligned} 
\end{equation}

\subsection{Formal Security Verification Simulation Based on AVISPA}

\begin{figure}[ht] 
	\centering  
	\fbox{\includegraphics[width=0.4\textwidth]{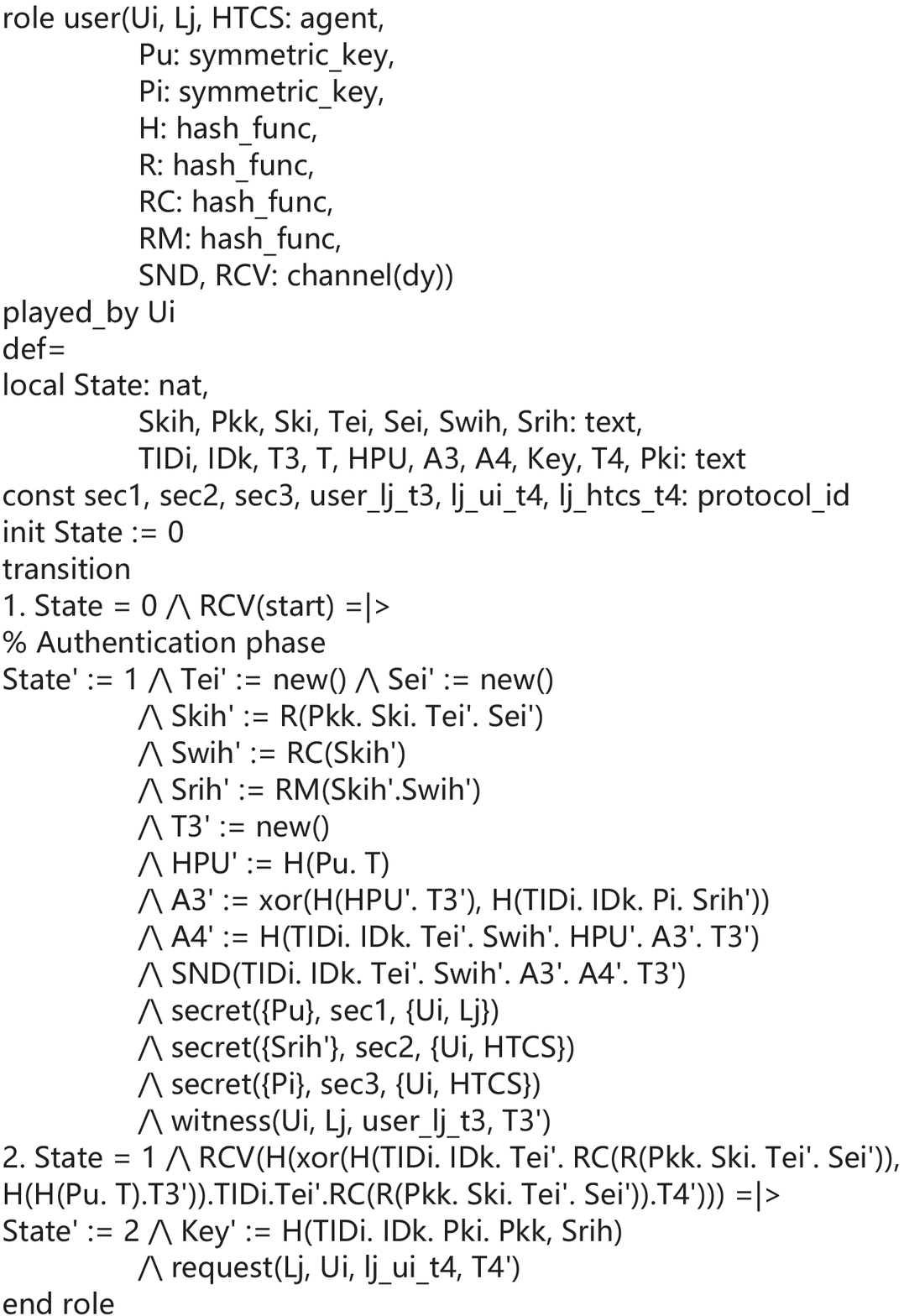}} 
	\caption{Role specification for user}  
	\label{Rolespecificationforuser}
\end{figure}

AVISPA \cite{AVISPA_intr} is a simulation verification tool for large-scale Internet security-sensitive protocols and has been widely used to construct defect-free authentication protocols, such as: \cite{Ostad-Sharif,ChallaECC-based,Dasavispa}, etc. So far, AVISPA supports a total of four security verification back-ends: On-the-fly Model-Checker (OFMC), Constraint Logic based Attack Searcher (CL-AtSe), SAT-based Model-Checker
(SATMC) and Tree Automata based on Automatic Approximations for the
Analysis of Security Protocols (TA4SP). Most scholars usually use OFMC and CL-AtSe back-ends to analyze the security of the network protocol. AVISPA outputs the protocol verification result through the following steps. First, use the language called High Level Protocol Specification Language (HLPSL) provided by AVISPA to represent the specifications of all roles and clarify their potential security attributes. Second, an automated translator called hlpsl2if to convert role specifications represented by HLPSL to the intermediate format (IF). For each back-end, IF is a lower-level and more acceptable format than HLPSL. Finally, 
role specifications in IF format is input to a back-end and output format (OF) is generated, where the result of OF can indicate whether the protocol is safe or unsafe.


\begin{figure}[ht] 
	\centering  
	\fbox{\includegraphics[width=0.4\textwidth]{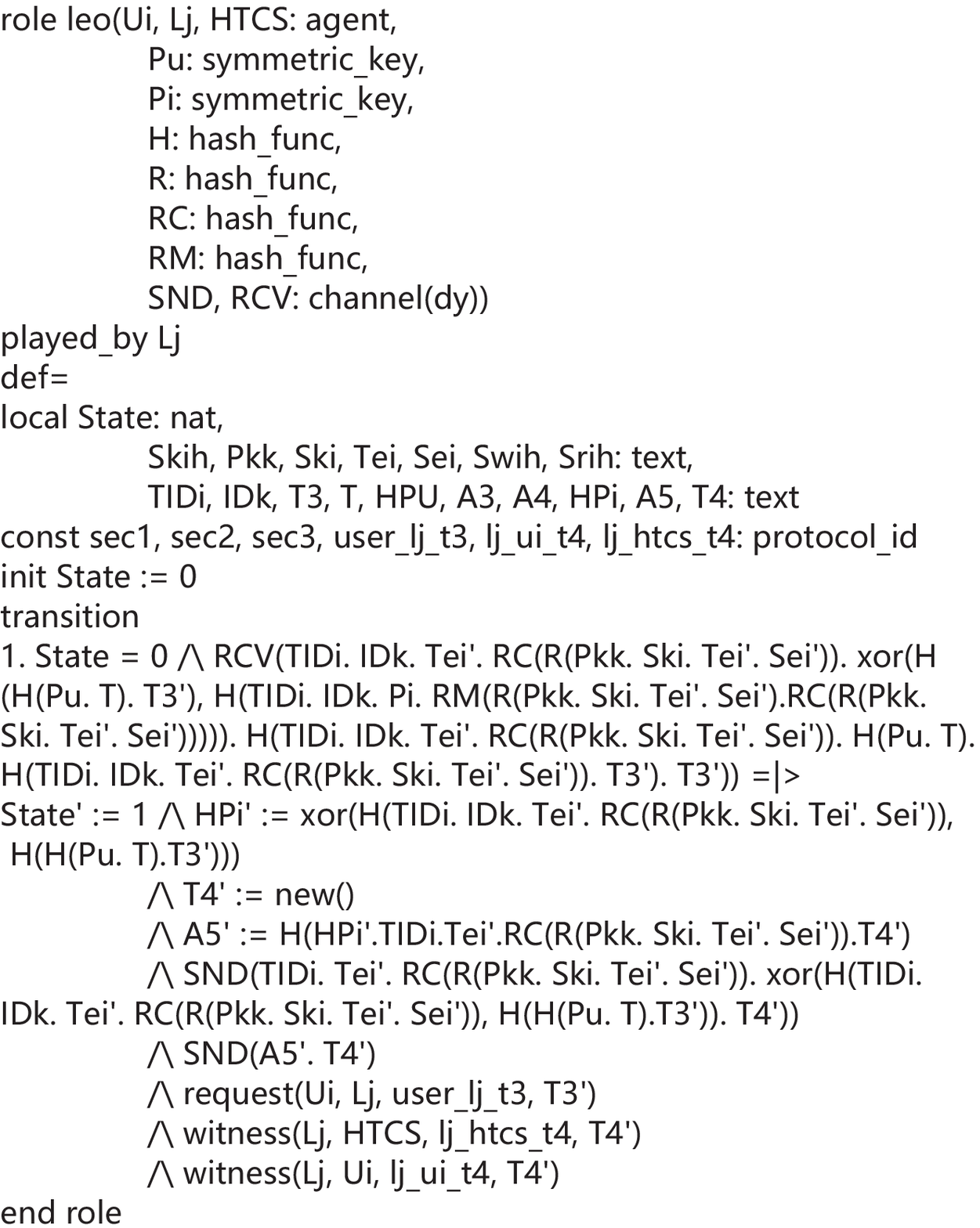}} 
	\caption{Role specification for satellite}  
	\label{Rolespecificationforsatellite}
\end{figure}


\begin{figure}[ht] 
	\centering  
	\fbox{\includegraphics[width=0.4\textwidth]{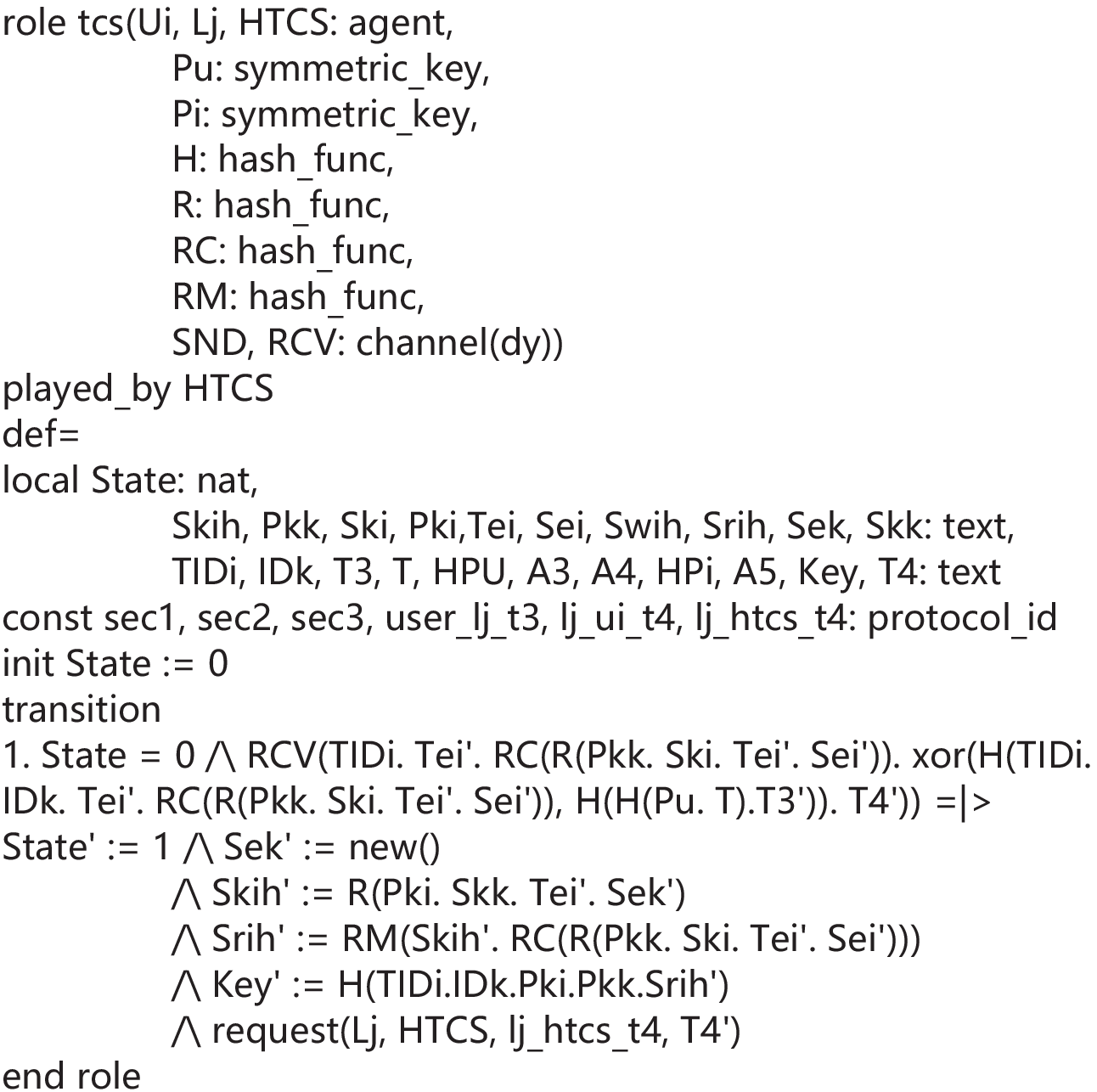}} 
	\caption{Role specification for TCS}  
	\label{RolespecificationforTCS}
\end{figure}


\begin{figure}[ht] 
	\centering  
	\fbox{\includegraphics[width=0.4\textwidth]{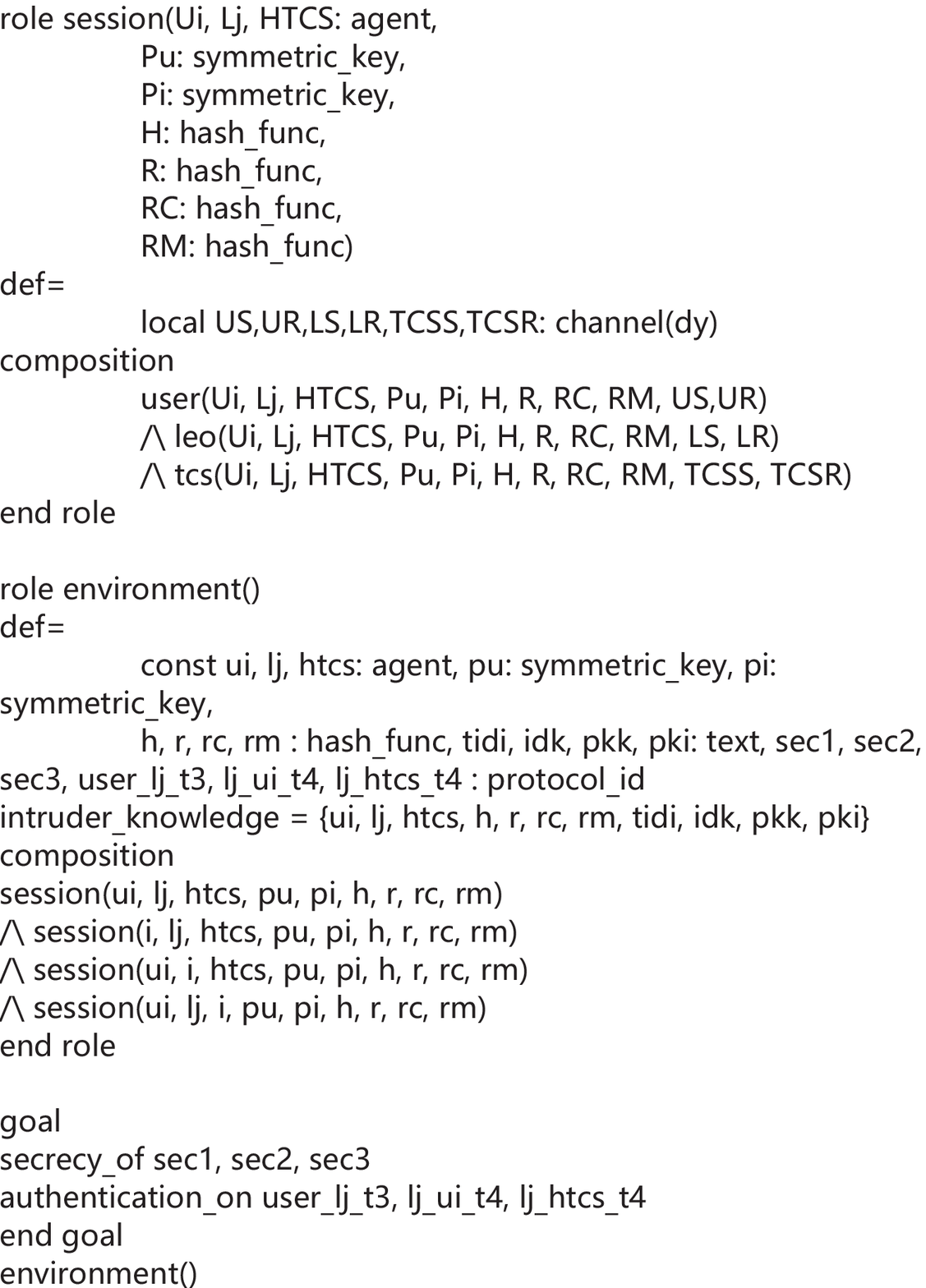}} 
	\caption{Role specification for session, environment and goal }  
	\label{Rolespecificationforother}
\end{figure}

\begin{figure}[ht] 
	\centering  
	\fbox{\includegraphics[width=0.25\textwidth]{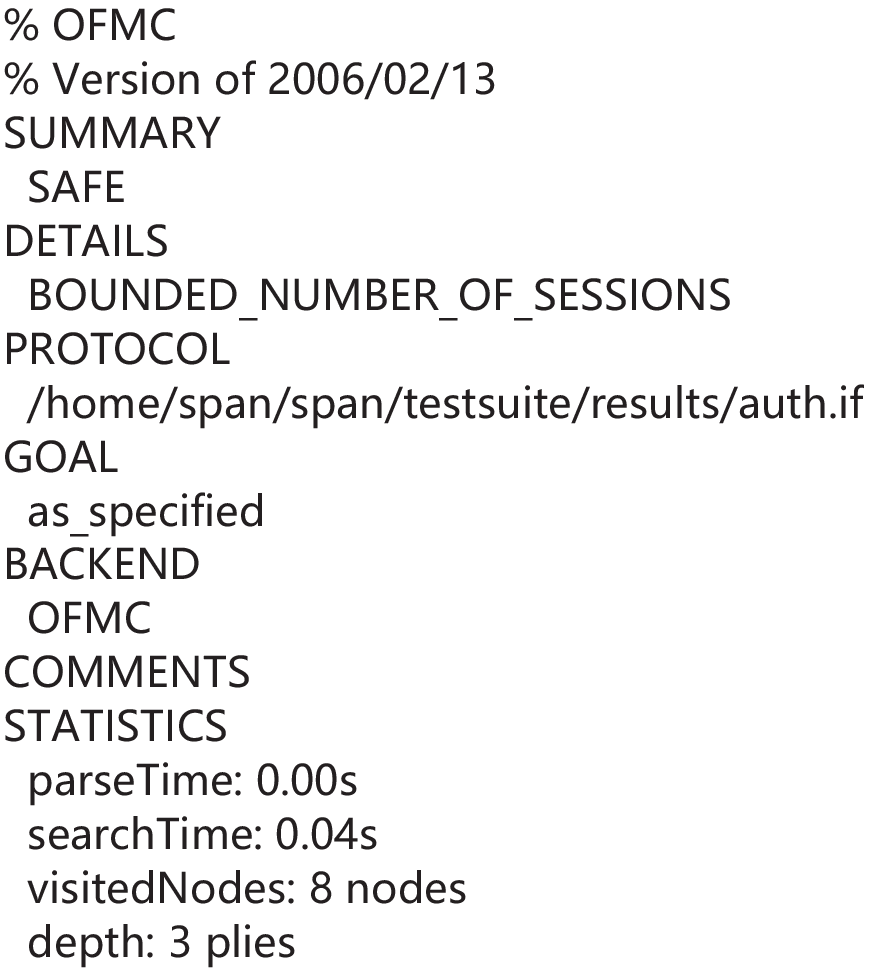}} 
	\caption{Simulation result using OFMC backend}  
	\label{SimulationresultusingOFMCbackend}
\end{figure}

\begin{figure}[ht] 
	\centering  
	\fbox{\includegraphics[width=0.25\textwidth]{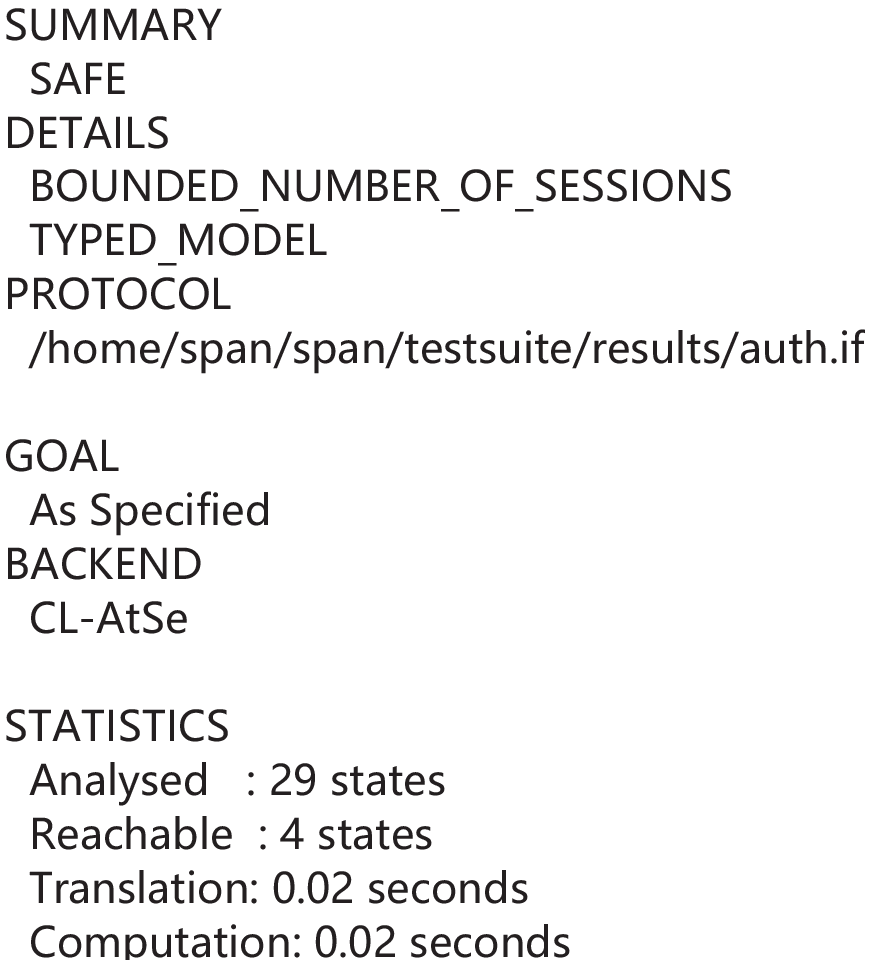}} 
	\caption{Simulation result using CL-AtSe backend}  
	\label{SimulationresultusingCLAtsebackend}
\end{figure}

Figures \ref{Rolespecificationforuser}, \ref{Rolespecificationforsatellite} and \ref{RolespecificationforTCS} respectively show the role specifications of the user, satellite node and TCS in HLPSL. In addition, the session, environment and goal are defined as shown in Figure \ref{Rolespecificationforother}, where the adversary is represented by i and can obtain all transmitted authentication messages based on Dolev-Yao model. Finally, we verify the security of PSAA protocol based on OFMC and CL-AtSe back-ends, the results are shown in Figures \ref{SimulationresultusingOFMCbackend} and \ref{SimulationresultusingCLAtsebackend} respectively. The simulation results of the two back-ends show that our proposed PSAA protocol is safe and does not have any attack path to enable the adversary to implement eavesdropping attacks, replay attacks and man-in-the-middle attacks.


\subsection{Informal Security Analysis}

\subsubsection{Mutual Authentication} \label{MutualAuthentication}
In the second step of the authentication phase, after receiving the access request $\{TID_i,ID_{tcs},te^{1}_{i-tcs},sw^{1}_{i-tcs},a_3,a_4,t_3\}$ from $u_i$, $L_j$ calculates $HPU^{'}_i=h(h(TID_i,HPU_j),t_3)$ and checks whether $a_4=h(TID_i,ID_{tcs},te^{1}_{i-tcs},sw^{1}_{i-tcs},HPU^{'}_i,a_3,t_3)$ is holds. $HPU_i$ can only be calculated by a user who has registered with the TCS, that is, $u_i$ is a legitimate user and the access request has not been modified if the check is passed. Similarly, after receiving the access response $\{a_5,t_4\}$ from $L_j$, $u_i$ confirms the legitimacy of $L_j$ by checking whether $a_5=h(HP_i,TID_i,te^{1}_{i-tcs},sw^{1}_{i-tcs},t_4)$ is established, because only $L_j$ after mutual authentication with TCS in the pre-negotiation phase can obtain the correct  authentication parameters $HP_i$ through $a_3\oplus HPU_i$. Therefore, the proposed PSAA protocol can meet the security requirements of mutual authentication.

\subsubsection{Key Negotiation} \label{KeyNegotiation}
In the third and fourth steps of the authentication phase, TCS and $u_i$ respectively calculate the session key $key=h(TID_i,ID_{tcs},pk_i,pk_{tcs},s\sigma^{1}_{i-tcs})$, where $s\sigma^{1}_{i-tcs}=Mod_2(sk^{1}_{i-tcs},sw^{1}_{i-tcs})$,  $sk^{1}_{i-tcs}=(pk_{tcs}\cdot sk_i+2\cdot te^{1}_{i-tcs})\cdot te^{1}_{i-tcs}+2\cdot re^{1}_{i-tcs}$ or $(pk_i\cdot sk_{tcs}+2\cdot te^{1}_{i-tcs}) \cdot te^{1}_{i-tcs} +2\cdot re^{1'}_{tcs}$. The session key is not determined by $u_i$ or TCS alone, but by the joint negotiation of the master public-private key pair of both parties. Moreover, the adversary cannot extract the master private keys of both parties through the eavesdropped $pk_i$, $pk_{tcs}$, $sw^{1}_{i-tcs}$ and other parameters in the public channel unless the adversary can solve the RLWE assumption in polynomial time. Therefore, the proposed PSAA protocol can meet the security requirements of key negotiation.

\subsubsection{Identity anonymous}
During the entire authentication phase, $u_i$’s true identity $ID_i$ is not transmitted in the public channel but the temporary identity $TID_i$, where $TID_i=ID_i\oplus s\sigma^{0}_{i-tcs}$. If the adversary attempts to obtain $ID_i$ through $TID_i \oplus s\sigma^{0}_{i-tcs}$, it can only compute $s\sigma^{0}_{i-tcs}$ from $pk_i$ and $pk_{tcs}$ by solving the RLWE assumption in  polynomial time, which is difficult for the adversary. Therefore, the adversary cannot obtain the true identity of the user and the proposed PSAA protocol can meet the security requirements of identity anonymous.


\subsubsection{Perfect forward/backward secrecy}
Each time when authenticating and negotiating a session key $key=h(TID_i,ID_{tcs},pk_i,pk_{tcs},s\sigma^{1}_{i-tcs})$, $u_i$ needs to randomly select $te^{1}_{i-tcs}$, $re^{1}_{i-tcs}$ from $\chi_{\beta}$ to calculate $sk^{1}_{i-tcs}$, $sw^{1}_{i-tcs}$ and $s\sigma^{1}_{i-tcs}$. Due to the randomness of $te^{1}_{i-tcs}$ and $re^{1}_{i-tcs}$, the session keys of the previous and the next session are not associated with the current session key, that is, even if the adversary obtains the current session key, it will not compromise the security of the forward and backward session keys. Therefore, the proposed PSAA protocol can meet the security requirements of perfect forward/backward secrecy.

\subsubsection{Eavesdropping attack} 
Similar as described in section \ref{KeyNegotiation}, the adversary can only obtain the parameters $\{TID_i,pk_i,ID_{tcs},pk_{tcs},te^{1}_{i-tcs},sw^{1}_{i-tcs},a_3,a_4,a_5,t_3,t_4\}$ transmitted in the public channel after an eavesdropping attack. The adversary cannot calculate the secret $s\sigma^{1}_{i-tcs}$ from the obtained $pk_i$, $pk_{tcs}$ and other parameters, unless the RLWE assumption is resolved in polynomial time. Therefore, the proposed PSAA protocol is resistant to eavesdropping attack.

\subsubsection{Replay attack}
All interactive messages in the authentication phase include a timestamp and a hash check value. If the adversary replays the eavesdropped original interactive information, all legal entities can simply compare the current timestamp with the timestamp of the message to determine whether it is maliciously replayed. Even if the adversary tampered with the timestamp in the replayed message, the adversary cannot modify the hash value due to the one-way feature of the hash function, and finally the malicious replayed message cannot be received by the legal entity. Therefore, the proposed PSAA protocol is resistant to replay attack.

\subsubsection{Impersonation attack}
The adversary can impersonate a user or a satellite node to launch impersonation attacks. When the adversary impersonates $u_i$ to send an access request to $L_j$, the adversary needs to obtain the parameter $pu_i$ that TCS distributes to the legitimate user $u_i$ during the registration phase. However, the adversary cannot extract $pu_i$ from the public parameters,
so a legitimate access request cannot be constructed to impersonate $u_i$. Similarly, when the adversary impersonates $L_j$, it is impossible to construct a legitimate response $\{a_5,t_4\}$ that passes user verification without obtaining $HPU_j$, where $a_5=h(HP^{'}_i,TID_i,te^{1}_{i-tcs},sw^{1}_{i-tcs},t_4)$ and $HP^{'}_i=a_3\oplus h(h(TID_i,HPU_j),t_3)$. Therefore, the proposed PSAA protocol is resistant to impersonation attack.

\subsubsection{Man-in-the-middle attack}
The adversary to implement the man-in-the-middle attack must be able to respectively impersonate $u_i$ to communicate with the $L_j$ and impersonate $L_j$ to communicate with $u_i$. However, in the proof of impersonation attack, it has been proven that the adversary cannot impersonate a legitimate user or satellite node. Therefore, the proposed PSAA protocol is resistant to man-in-the-middle attack.

\subsubsection{Device loss attack}
When the user's device is lost and acquired by the adversary, the adversary cannot log in to the device and access SIN as a legitimate node because the user's true identity $ID_i$, password $PW_i$ and biometric $BIO_i$ cannot be obtained. Furthermore, even if the adversary obtains $\{DID_i,DP_i,DPU_i,DSK_i,ver_i,v_i\}$ stored in the device through power analysis, it is impossible to calculate authentication parameters such as the user's master private key $sk_i$ without $ID_i$, $PW_i$ and $BIO_i$. Therefore, the proposed PSAA protocol is resistant to device loss attack.

\subsubsection{Insider attack}
The insider attack refers to a privileged insider in TCS attempting to obtain the user's password $PW_i$, biometric $BIO_i$, and the master private key $sk_i$ during user registration. However, the registration request submitted by $u_i$ does not directly contain $PW_i$, $BIO_i$ and $sk_i$, but $RPW_i$ and $pk_i$, where $RPW_i=h(PW_i,\sigma_i)$, $pk_i=a\cdot sk_i + 2\cdot se_i$. Due to the one-way nature of the hash function and the difficult assumption of RLWE, the adversary cannot get $PW_i$, $BIO_i$ and $sk_i$ from $RPW_i$ and $pk_i$. Therefore, the proposed PSAA protocol is resistant to insider attack.

\subsubsection{Quantum attack}
The session key negotiation mechanism of the PSAA protocol is designed based on the RLWE assumption, in which RLWE has been proved to be an efficient and secure post-quantum cryptography primitive and can be reduced to SVP in the lattice. In addition, we refer to the randomization mechanism proposed in \cite{Gaorandomized} to add additional unpredictable random values to avoid public-private key pair reuse attack. Therefore, the proposed PSAA protocol is resistant to quantum attack.

\section{Performance Comparison} \label{PerformanceComparison}
In this section, we compare the proposed PSAA protocol with the other five latest related protocols \cite{Ostad-Sharif,QiECCsatellite,AnFRA,Xuehandover,ma2019laa} in SIN in terms of security attributes, authentication delay, and communication overhead. In order not to lose generality, we use the NTL library \cite{NTL,DingKeyexchange} and Openssl library to simulate the cryptographic primitives of these protocols in  C language without multithreading and parallelism, then run on a computer with Intel(R) Core(TM) i7-7820HQ CPU @ 2.90GHz processor and 8.00 GB RAM. At the same time, to ensure the security strength of the protocol, we set parameters $n$, $q$, $\beta$ to 1024, 120833, 2.6, respectively, and the parameters selected make the security strength of PSAA sufficient to cover the security strength of AES-192 and AES-256 \cite{DingKeyexchange}. In addition, for other related protocols, set the elliptic curve to secp256r1, the output length of the hash function $h$ is 256 bits, the length of the temporary identity $TID_i$ and the timestamp is 100 bits, the size of the point in the additive cyclic group is 512 bits, the symmetric encryption/decryption algorithm is AES-256. Moreover, we ensure the accuracy of the time cost of each cryptographic primitive by operating 10,000 times on the relevant cryptographic primitives and taking the average value.

\subsection{Comparison of Security Attributes}

\begin{table}[htp]
	\caption{Comparison of security attributes}
	\centering
	\resizebox{0.5\textwidth}{!}{
	\begin{tabular}{p{3.2cm} <{}  p{0.55cm} <{\centering} p{0.55cm} <{\centering} p{0.55cm} <{\centering} p{0.55cm} <{\centering} p{0.55cm} <{\centering} p{0.5cm} <{\centering} }
		\hline
		\specialrule{0em}{1pt}{1pt}
		Security attributes	&	\cite{Ostad-Sharif}	&	\cite{QiECCsatellite}	&	\cite{AnFRA}	&	\cite{Xuehandover}	& \cite{ma2019laa}	&	Our	\\
		\hline
		\specialrule{0em}{1pt}{1pt}
		Mutual authentication	&	\checkmark	&	\checkmark	&	\checkmark	&	\checkmark&\checkmark &	\checkmark\\ \specialrule{0em}{1pt}{1pt}
		\specialrule{0em}{1pt}{1pt}
		Key negotiation			&	\checkmark	&	\checkmark	&	\checkmark	&	\checkmark&\checkmark &	\checkmark\\ \specialrule{0em}{1pt}{1pt}
		\specialrule{0em}{1pt}{1pt}
		Identity anonymous		&	\checkmark	&	\checkmark	&	\checkmark	&	\checkmark&\checkmark &	\checkmark\\ \specialrule{0em}{1pt}{1pt}
		\specialrule{0em}{1pt}{1pt}
		PFS/FBS  &	\checkmark	& \checkmark & \checkmark	&	\checkmark & \checkmark &	\checkmark\\ \specialrule{0em}{1pt}{1pt}
		\specialrule{0em}{1pt}{1pt}
		Eavesdropping attack	&	\checkmark	&	\checkmark	& \checkmark &	\checkmark & \checkmark &\checkmark	\\ \specialrule{0em}{1pt}{1pt}
		\specialrule{0em}{1pt}{1pt}
		Replay attack	&	\checkmark	&	\checkmark	& \checkmark &	\checkmark & $\times$ &\checkmark	\\ \specialrule{0em}{1pt}{1pt}
		\specialrule{0em}{1pt}{1pt}
		Impersonation attack	&	$\times$	&	$\times$	& \checkmark &	\checkmark	& \checkmark &	\checkmark\\ \specialrule{0em}{1pt}{1pt}
		\specialrule{0em}{1pt}{1pt}
		Man-in-the-middle attack	&	\checkmark	&	\checkmark	& \checkmark &	\checkmark & \checkmark&\checkmark	\\ \specialrule{0em}{1pt}{1pt}
		\specialrule{0em}{1pt}{1pt}
		Device loss attack	&	\checkmark	&	$\times$	&	$\times$ &	$\times$ & $\times$ &	\checkmark	\\ \specialrule{0em}{1pt}{1pt}
		\specialrule{0em}{1pt}{1pt}
		Insider attack	&	\checkmark	& \checkmark & \checkmark  & \checkmark & \checkmark&	\checkmark	\\ \specialrule{0em}{1pt}{1pt}
		\specialrule{0em}{1pt}{1pt}
		Quantum attack	&	$\times$	& $\times$	& $\times$ & $\times$ &\checkmark &	\checkmark \\ \specialrule{0em}{1pt}{1pt}
		\hline       
	\end{tabular}
}
	\label{Comparison of security attributes}
\end{table}

As shown in Table \ref{Comparison of security attributes}, we evaluate and compare the security attributes of related protocols and our proposed PSAA protocols for mutual authentication, key negotiation, identity anonymity, perfect forward/backward secrecy (PFS/FBS), and the ability to withstand eavesdropping, replay, impersonation, man-in-the-middle, device loss, insider and quantum attacks.
Among them, \cite{Ostad-Sharif} and \cite{QiECCsatellite} do not check the validity of the satellite node during the operation of the entire protocol, which will be vulnerable to satellite impersonation attacks. For protocols \cite{AnFRA} and \cite{Xuehandover}, due to the lack of a login verification mechanism, an adversary can use the user’s lost legitimate device as a legitimate node to access SIN. In addition, even though \cite{QiECCsatellite} has a simple login mechanism, it only verifies the user password, which will also make the protocol vulnerable to offline password guessing attacks. The security of all four related protocols is based on the difficult problem of elliptic curve cryptography, which has been proven to be breakable in polynomial time in the post-quantum era, so these protocols are vulnerable to quantum attacks. 
Besides, although the lattice-based protocol \cite{ma2019laa} can resist quantum attacks, it lacks detailed mechanisms for historical authentication message replay verification and user three-factor login verification, so it cannot resist replay attacks and device loss attacks.
Therefore, from the security analysis and the security attribute comparison results in Table \ref{Comparison of security attributes}, it is obvious that our proposed PSAA protocol can meet all security attributes and its security is better than other protocols.

\subsection{Comparison of Authentication Delay}

\begin{table}[htp]
	\caption{Execution time of various primitive operations}
	\centering
	\resizebox{0.5\textwidth}{!}{
	\begin{tabular}{p{1 cm} <{} p{4.9 cm} <{} p{1.2 cm} <{}  }
		\hline
		\specialrule{0em}{1pt}{1pt}
		Notation	&	Description		&Time (us)	\\
		\hline
		\specialrule{0em}{1pt}{1pt}
		$T_{h}$		&	Hash operation	&	0.372	\\ \specialrule{0em}{1pt}{1pt}
		\specialrule{0em}{1pt}{1pt}
		$T_{bp}$	&	Pairing operation&	624.168	\\ \specialrule{0em}{1pt}{1pt}
		\specialrule{0em}{1pt}{1pt}
		$T_{ed}$	&	Encryption/decryption operation&6.028	\\ \specialrule{0em}{1pt}{1pt}
		\specialrule{0em}{1pt}{1pt}
		$T_{mul}$	&	 ECC point multiplication operation&10.67	\\ \specialrule{0em}{1pt}{1pt}
		\specialrule{0em}{1pt}{1pt}
		$T_{add}$	&	ECC point addition operation&	0.711	\\ \specialrule{0em}{1pt}{1pt}
		\specialrule{0em}{1pt}{1pt}
		$T_{samp}$	&	Sampling operation on $\chi_{\beta}$ &  25.763	\\ \specialrule{0em}{1pt}{1pt}
		\specialrule{0em}{1pt}{1pt}
		$T_{rmul}$	&	Multiplication operation in $R_q$	&  202.786	\\ \specialrule{0em}{1pt}{1pt}
		\specialrule{0em}{1pt}{1pt}
		$T_{radd}$	&	Addition operation in $R_q$	&	11.702	\\ \specialrule{0em}{1pt}{1pt}
		\specialrule{0em}{1pt}{1pt}
		$T_{cha}$	&	$Cha$ operation &17.002	\\ \specialrule{0em}{1pt}{1pt} 
		\specialrule{0em}{1pt}{1pt}
		$T_{lmul}$	&	Matrix-vector multiplication operation & 1025.011 \\ \specialrule{0em}{1pt}{1pt} 
		\specialrule{0em}{1pt}{1pt}
		$T_{ladd}$	&  	Matrix addition operation 	& 1060.936	\\ \specialrule{0em}{1pt}{1pt} 
		
		\hline       
	\end{tabular}
	}
	\label{Execution time of various primitive operations}
\end{table}

The authentication delay refers to the time interval from when a user initiates an access request to when the session key is negotiated with TCS in the authentication phase, which includes the computational overhead and message transmission time.  Figure \ref{Execution time of various primitive operations} shows the each execution time of the relevant cryptographic primitives involved in the protocols we obtained through experiments, and the computational overhead of $Mod_2$ is small enough to be neglected as \cite{QiFeng}. Moreover, the transmission time $T_{u-s}$ of the authentication message in the satellite-ground link is reasonably set to 10 ms \cite{Xuehandover}. It is worth mentioning that  when analyzing the computational overhead, we do not consider the calculation time required in the user login verification phase.

\begin{table}[htp]
	\caption{Comparison of authentication delay}
	\centering
	\resizebox{0.5\textwidth}{!}{
	\begin{tabular}{p{1cm} <{}  p{5cm} <{} p{1.4cm} <{\centering} p{1.6cm} <{\centering}}
		\hline
		\specialrule{0em}{1pt}{1pt}
		Protocol	&	Computational overhead  & Transmission time & Total	\\
		\hline
		\specialrule{0em}{1pt}{1pt}
		\cite{Ostad-Sharif}	&	$6T_{mul}+9T_{h}$	& $4T_{u-s}$	&40.067 ms	\\ \specialrule{0em}{1pt}{1pt}
		\specialrule{0em}{1pt}{1pt}
		\cite{QiECCsatellite}	&	$6T_{mul}+10T_{h}+2T_{ed}$ & $6T_{u-s}$	&60.080 ms \\ \specialrule{0em}{1pt}{1pt}
		\specialrule{0em}{1pt}{1pt}
		\cite{AnFRA}	&	$\approx 14T_{mul}+2T_{bp}+3T_{h}$ & $2T_{u-s}$ & 21.400 ms	\\ \specialrule{0em}{1pt}{1pt}
		\specialrule{0em}{1pt}{1pt}
		\cite{Xuehandover}	&	$9T_{mul}+6T_{h}+4T_{add}+2T_{ed}$& $2T_{u-s}$ & 20.113 ms	\\ \specialrule{0em}{1pt}{1pt}
		\specialrule{0em}{1pt}{1pt}
		\cite{ma2019laa}	&	$5T_{lmul}+3T_{ladd}+7T_{h}+2T_{ed}$ & $2T_{u-s}$ & 28.322 ms	\\ \specialrule{0em}{1pt}{1pt}
		\specialrule{0em}{1pt}{1pt}
		Our	&	$3T_{samp}+T_{cha}+8T_{rmul}+4T_{radd}+11T_{h}+2T_{ed}$ & $2T_{u-s}$ & 21.780 ms	\\ \specialrule{0em}{1pt}{1pt}

		\hline       
	\end{tabular}
	}
	\label{Comparison of Authentication Delay}
\end{table}


For \cite{Ostad-Sharif}, the computational overhead for the entire authentication phase includes six ECC point multiplication operations and nine hash operations, which is equal to $6T_{mul}+9T_{h}=0.067$ ms, and the transmission time required for authentication messages to be transmitted four times sequentially in the satellite-ground link is $4T_{u-s}=40$ ms, so the total authentication delay is 40.067 ms.

Similarly for the \cite{QiECCsatellite}, the entire computational overhead includes six ECC point multiplication operations, ten hash operations and two encryption/decryption operations, which is equal to $6T_{mul}+10T_{h}+2T_{ed}=0.080$ ms, and the transmission time required for authentication messages is $6T_{u-s}=60$ ms, so the total authentication delay is 60.080 ms. 

For \cite{AnFRA}, the entire computational overhead approximately include fourteen ECC point multiplication operations, two pairing operations and three hash operations, which is equal to $14T_{mul}+2T_{bp}+3T_{h}=0.400$ ms, and the transmission time is only $2T_{u-s}=20$ ms due to the parallel transmission of messages in the authentication phase, so the total authentication delay is 20.400 ms. 

For \cite{Xuehandover}, the entire computational overhead includes nine ECC point multiplication operations, six hash operations, four ECC point addition operations and two encryption/decryption operations, which is equal to $9T_{mul}+6T_{h}+4T_{add}+2T_{ed}= 0.113$ ms, and the parallel transmission of messages like \cite{AnFRA} causes the authentication message transmission time to be $2T_{u-s}=20$ ms, so the total authentication delay is 20.113 ms. 

For \cite{ma2019laa}, the entire computational overhead includes five matrix-vector multiplication operations, three matrix addition operations, seven hash operations and two encryption/decryption operations, which is equal to $5T_{lmul}+3T_{ladd}+7T_{h}+2T_{ed}=8.322$ ms, and the transmission time is $2T_{u-s}=20$ ms, so the total authentication delay of \cite{ma2019laa} is 28.322 ms. 

For the PSAA protocol we proposed, the computational overhead of user, satellite, and TCS are $2T_{samp}+T_{cha}+4T_{rmul}+2T_{radd}+5T_{h}$, $4T_{h}+T_{ed}$, and $T_{samp}+4T_{rmul}+2T_{radd}+2T_{h}+T_{ed}$, respectively, and the total computational overhead is $3T_{samp}+T_{cha}+8T_{rmul}+4T_{radd}+11T_{h}+2T_{ed}=1.780$ ms. Moreover, in the second step of the authentication phase, the satellite sends messages to both the user and the TCS at the same time, resulting in the total transmission time is $2T_{u-s}=20$ ms. Therefore, the authentication delay of PSAA protocol is 21.780 ms. 

From Table \ref{Comparison of Authentication Delay}, our proposed PSAA protocol is far superior to \cite{Ostad-Sharif} and \cite{QiECCsatellite} in terms of authentication delay and reduces the delay of about 18.287 ms and 38.3 ms, respectively, which greatly improves the quality of service for users in SIN. In addition, the PSAA protocol reduces the authentication delay of 6.542 ms compared with the same anti-quantum protocol \cite{ma2019laa}, which is more efficient.
Besides, although the authentication delay of \cite{AnFRA} and \cite{Xuehandover} is slightly lower than that of the PSAA protocol we proposed, the PSAA protocol in the post-quantum era provides higher security and anti-quantum properties, which can also better serve users.

\subsection{Comparison of Communication Overhead}

\begin{table}[htp]
	\caption{Comparison of communication overhead}
	\centering
	\begin{tabular}{p{1.2cm} <{}  p{1.4cm} <{} p{1.6cm} <{} p{1.6cm} <{} p{1.6cm} <{}}
		\hline
		\specialrule{0em}{1pt}{1pt}
		Protocol	&	User  & Satellite & TCS & Total	\\
		\hline
		\specialrule{0em}{1pt}{1pt}
		\cite{Ostad-Sharif}	& 968 bits	&  1836 bits	&868 bits& 3672 bits	\\ \specialrule{0em}{1pt}{1pt}
		\specialrule{0em}{1pt}{1pt}
		\cite{QiECCsatellite}	& 1580 bits	 &  2648 bits &968 bits	&  5196 bits\\ \specialrule{0em}{1pt}{1pt}
		\specialrule{0em}{1pt}{1pt}
		\cite{AnFRA}	& 1168 bits &  3360 bits  &  - & 4528 bits	\\ \specialrule{0em}{1pt}{1pt}
		\specialrule{0em}{1pt}{1pt}
		\cite{Xuehandover}	& 2392 bits  &  2804 bits & - & 5196 bits	\\ \specialrule{0em}{1pt}{1pt}
		\specialrule{0em}{1pt}{1pt}
		\cite{ma2019laa}	&  0.781 MB  &  400.430 MB & - &  401.211 MB	\\ \specialrule{0em}{1pt}{1pt}
		\specialrule{0em}{1pt}{1pt}
		Our	& 19244 bits & 19244 bits  & -& 38488 bits	\\ \specialrule{0em}{1pt}{1pt}
		
		\hline       
	\end{tabular}
	\label{Comparison of Communication Overhead}
\end{table}

The total size of authentication messages sent or forwarded by all entities in the authentication phase is considered as the communication overhead. 

For \cite{Ostad-Sharif}, the authentication messages transmitted by user, satellite and TCS are $\{TID_i,E_i,S_i,T_i\}$, $\{TID_i,E_i,S_i,T_i,ID_{leo},X_{tcs},S_{tcs}\}$ and $\{TID_i,X_{tcs},S_{tcs}\}$ respectively, where $TID_i=ID_i\oplus h(E_i||F_i)$, $\{E_i,X_{tcs}\}$ are the points of the elliptic curve, $\{S_i,S_{tcs}\}$ are hash values, $T_i$ is the timestamp, $ID_{leo}$ is the identity. Therefore, the communication overhead of user, satellite, and tcs are $100+512+256+100=968$ bits, $100+512+256+100+100+512+256=1836$ bits, and $512+256+100=868$ bits, respectively, and the total communication overhead of \cite{Ostad-Sharif} is 3672 bits.

For \cite{QiECCsatellite}, the authentication messages transmitted by user, satellite and TCS are $\{c,X,t_1,Z\}$, $\{c,X,t_1,ID_{leo},A_{tcs},Y,t_2,Z,ID_{leo}\}$ and $\{A_{tcs},Y,t_2,ID_{leo}\}$ respectively, where $c$ is the symmetrically encrypted ciphertext of $\{ID_i,R_i,H(ID_i,R_i),t_1\}$, $\{X,Y\}$ are the points of the elliptic curve, $\{t_1, t_2\}$ are timestamps, $\{ID_{leo},ID_{i}\}$ are identities, $\{R_i,A_{tcs},Z\}$ are hash values. Therefore, the communication overhead of user, satellite, and tcs are $100+256+256+100+512+100+256=1580$ bits, $100+256+256+100+512+100+100+256+512+100+256+100=2648$ bits, $256+512+100+100=968$ bits, respectively, and the total communication overhead of \cite{QiECCsatellite} is 5196 bits.

For \cite{AnFRA}, the authentication messages transmitted by user, satellite and TCS are $\{TID_{u},ID_{fleo},ID_{tcs},g^{r_{u}},ts_3,\sigma_{u}\}$, $2\{TID_{u},ID_{fleo},ID_{ftcs},g^{r_u},g^{r_{ftcs}},ts_4,\sigma_{ftcs}\}$, $\varnothing$  respectively, where $\{TID_{u},ID_{fleo},ID_{tcs}\}$ are identities, $\{g^{r_{u}},g^{r_{ftcs}}\}$ are points in the additive cyclic group, $\{\sigma_{u},\sigma_{ftcs}\}$ are hash values, $\{ts_3, ts_4\}$ are timestamps. Therefore, the communication overhead of user, satellite are $100   +   100   +   100   +   512   +   100   +   256=1168$ bits, $2\times \{100   +   100   +   100   +   512   +   512   +   100   +   256\} =3360 $ bits, respectively, and the total communication overhead of \cite{AnFRA} is 4528 bits. 

For \cite{Xuehandover}, the authentication messages transmitted by user, satellite and TCS are $\{ID_{leo},R_j,P_j,v_j,TS_1\}$, $\{2TID_{j},P_{leo},R_{tcs},v_{leo},R_{leo},2TS_2,R_j\}$, $\varnothing$  respectively, where $P_{j}=\{pk_j,TID_j,Enc(pk_{ncc},TID_j,ID_j),LT_j\}$, $P_{leo}=\{pk_{leo},ID_{leo}\}$, $\{ID_{leo},TID_j,ID_j\}$ are identities, $\{R_j,R_{leo},R_{tcs},pk_j,pk_{ncc},pk_{leo}\}$ are the points of the elliptic curve, $\{TS_1,TS_2\}$ are timestamps, the length of $v_j$ and $v_{leo}$ is 256bits. Therefore, the communication overhead of user, satellite are $100   +   512   +   512   +   100   +   512   +   100   +   100   +   100   +   256   +   100 =2392$ bits, $2\times100   +   512   +   100   +   512   +   256   +   512   +   2\times100   +   512 =2804$ bits, respectively, and the total communication overhead of \cite{Xuehandover} is 5196 bits.

For \cite{ma2019laa}, the authentication messages transmitted by user, satellite and TCS are $\{C_i,e_i\}$, $\{MAC_s,U_G,M_S,Enc(C_i,M_i)\}$, $\varnothing$  respectively, where $MAC_s$ is hash value, $C_i\in Z^{m+1}_{q1}$, $e_i\in Z^{m}_{q1}$, $U_G\in Z^{n\times (m+1)}_{q1}$, $\{M_S,M_i\}\in Z^{m+1}_2$, and $q1=n^2= $ 1048576, $m=8n\log_2 q1 = 163840$ \cite{ma2019laa}, so the sizes of $C_i$, $e_i$, $U_G$, $M_S$, $M_i$ are 3276820 bits, 3276800 bits, 400 MB, 163841 bits, 163841 bits respectively. 
Therefore, the communication overhead of user, satellite are 0.781 MB, 400.430 MB, the total communication overhead of \cite{ma2019laa} is 401.211 MB. 

According to \cite{GaoPostQuantum}, the size of the element in $\chi_{\beta}$ is $n\cdot \lceil \log_2 q  \rceil$ bits, so the size of $te^{i}_{i-tcs}$ transmitted in the PSAA protocol is 17408 bits. In the authentication phase of the PSSLA protocol, a total of three messages $\{TID_i,ID_{tcs},te^{1}_{i-tcs},sw^{1}_{i-tcs},a_3,a_4,t_3\}$, $\{a_5,t_4\}$, and $\{TID_i,te^{1}_{i-tcs},sw^{1}_{i-tcs},HP_i,t_4\}$ are transmitted, and their sizes are $100+100+17408+1024+256+256+100=19244$ bits, $256+100=356$ bits, and $100+17408+1024+256+100=18888$ bits, respectively. Therefore, the total communication overhead of our proposed PSAA protocol is 38488 bits.


From the comparison of the communication overhead shown in Table \ref{Comparison of Communication Overhead}, it can be concluded that the communication overhead of PSAA is greater than that of protocols \cite{Ostad-Sharif,QiECCsatellite,AnFRA,Xuehandover}.
This is because the elements in RLWE, such as $te^{1}_{i-tcs}$, are much larger in size than the traditional cryptography mechanism, the common drawback of the quantum-resistant lattice-based cryptography mechanism \cite{gupta2020laac,Gaorandomized,GaoComparison,alkim2016newhope} result in the PSAA protocol has higher communication overhead. 
In addition, compared with the quantum-resistant protocol \cite{ma2019laa}, the PSAA protocol significantly reduces the communication overhead and makes SIN more available.
Although the PSAA protocol has disadvantages in communication overhead compared with non-anti-quantum protocols, it has higher security and quantum resistance with little increase in authentication delay, and with the development of satellite communication technology, satellites already have higher communication capabilities, so we believe that the proposed PSAA protocol is more suitable for SIN in the post-quantum era.

\section{Conclusion} \label{Conclusion}

In this paper, we first analyze the deficiencies and security defects of the existing related protocols, and summarize the security requirements of a well-designed protocol in SIN, and then propose a new PSAA protocol based on randomized RLWE, which can not only meets the security requirements of mutual authentication, key negotiation, identity anonymity and resist various known attacks, but also has quantum resistance. 
In the PSAA protocol, the satellite verifies the user’s legitimacy and allows the user to negotiate a session key with TCS, which reduces the authentication delay and guarantees the quality of service of the access user. 
For the proof of the security of the protocol, two formal analysis methods, ROM model and AVISPA, and the informal analysis method, demonstrate that the PSAA protocol meets all the security requirements previously summarized. Moreover, performance analysis and comparison show that the PSAA protocol is more suitable for SIN than other protocols under the premise of ensuring security and quantum resistance.

\ifCLASSOPTIONcompsoc
  \section*{Acknowledgments}
\else
  \section*{Acknowledgment}
\fi

The authors would like to thank all the anonymous reviewers for their helpful advice. 

\ifCLASSOPTIONcaptionsoff
  \newpage
\fi



%

\bibliography{IEEEabrv}

\end{document}